\documentclass[acmsmall,screen]{acmart}

\AtBeginDocument{%
  }
\newcommand{\re}[1]{\textcolor{black}{#1}}

\setcopyright{acmlicensed}
\copyrightyear{2026}
\acmYear{2026}
\acmDOI{XXXXXXX.XXXXXXX}

\acmConference[CSCW]{Conference on Computer-Supported Cooperative Work and Social Computing.}{October, 2026}{USA}
\acmISBN{978-1-4503-XXXX-X/18/06}

\usepackage{tabularx}
\usepackage{booktabs}
\usepackage{ragged2e}
\usepackage{enumitem}
\usepackage{xcolor}
\usepackage{array}
\usepackage{makecell} 

\setcopyright{cc}
\setcctype{by}

\copyrightyear{2026}
\acmYear{2026}
\acmDOI{10.1145/3816940}

\acmJournal{PACMHCI}
\acmVolume{10}
\acmNumber{6}
\acmArticle{CSCW092}
\acmMonth{10}

\begin{document}

\title[Challenges, Design Opportunities, and Concerns]{Generative AI as a Mediator in Creator Collaboration: Challenges, Design Opportunities and Concerns}

\author{Hajun Kim}
\authornote{These authors contributed equally to this work.}
\affiliation{%
  \institution{Independent Researcher}
  \city{Seoul}
  \country{Republic of Korea}
}
\email{hajun0628@gmail.com}

\author{Jini Kim}
\authornotemark[1]
\affiliation{%
  \institution{Carnegie Mellon University}
  \city{Pittsburgh}
  \state{PA}
  \country{USA}
}
\email{jinik@andrew.cmu.edu}

\author{Yunjae Josephine Choi}
\affiliation{%
  \institution{KAIST}
  \city{Daejeon}
  \country{Republic of Korea}
}
\email{choi.yj@kaist.ac.kr}


\begin{abstract}
The development of GenAI tools for collaboration among creators has recently attracted significant attention. Despite their growing use, prior studies have yet to fully explore the broader challenges of creative collaboration or the role of GenAI in supporting creators from diverse fields. To address this gap, we conducted a multi-method study with narrative writers (N = 10) and visual designers (N = 10) in the context of game development, where diverse collaboration challenges and asymmetries are especially salient. Our findings suggest that future GenAI tools have the potential to bridge the knowledge gap and workflow gaps, reduce the psychological burden of interacting with creators of different background, and improve work efficiency. At the same time, we highlight concerns that GenAI may reduce communication among creators, heighten reputational risks, and reshape interpersonal dynamics and collaborative processes. We also identify design opportunities to ensure that GenAI tools genuinely enhance collaboration without undermining the creativity and values inherent to human work.
\end{abstract}

\maketitle

\begin{figure}[h]
    \centering
    \includegraphics[width=\textwidth]{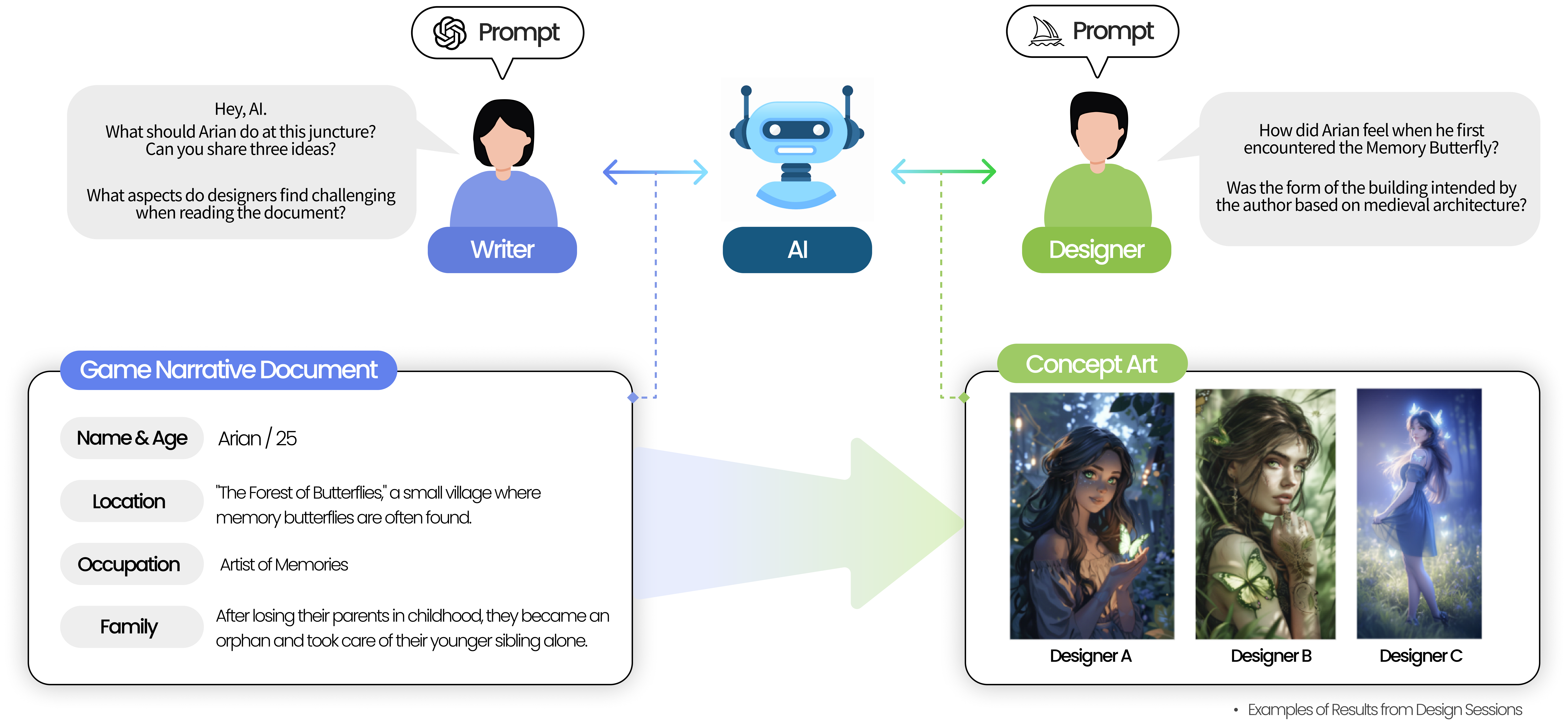}
    \caption{Overview of AI Collaboration Between Game Narrative Writers and Visual Designers}
    \label{fig:workflow}
\end{figure}

\section{\MakeUppercase{Introduction}}
Generative AI (GenAI) has begun to integrate into the workflows of individual content creators, innovating individual creative processes from ideation to the automation of repetitive tasks \cite{Maeda2022, doshi2023generative, sun2024generative, adobe2024generative, moruzzi2024user}. Moving beyond these individual-centric use cases, the focus of the HCI community has recently expanded toward integrating GenAI into multi-party collaborative processes \cite{han2024teams, zhou2024understanding}. Early studies have concentrated on exploring new collaborative dynamics by leveraging GenAI as a teammate, such as by coordinating its intervention timing or control \cite{wang2025social, houde2025controlling}. However, the discussion needs to deepen from treating GenAI merely as a team member to address the core challenge of creative collaboration long emphasized by CSCW researchers: building \textit{Common Ground} \cite{clark1991grounding, dourish1992awareness, fussell2000coordination, whittaker1998dynamics, kraut2002understanding}—that is, the body of mutual knowledge and beliefs that collaborators should continuously construct and update through an effortful process of communication.

\re{Building common ground becomes particularly challenging when heterogeneous creators convene \cite{muller2019learning}. Due to differences in their professional domains \cite{munster2014common, d2017comfy}, perspectives, and goals \cite{billinghurst1999asymmetries}, these creators possess epistemic asymmetry, which induces continuous friction in the process of building shared understanding \cite{gardiner2020learning, feast2012professional,nguyen2020multidisciplinary}. Furthermore, structural conditions including remote, asynchronous work environments \cite{koschmann2003reconsidering, olson2000distance, morrison2020challenges, kleinsmann2008barriers} or hierarchical dynamics \cite{gray2022confronting, grimm2024building} within a team create additional burdens in coordinating these frictions \cite{okhuysen200910}. Traditionally, organizations have mitigated these issues by utilizing boundary objects, such as managing joint documents through iterative feedback \cite{lee2007boundary, gonzalez2025achieving, saad1996shared}, or employing coordination mechanisms that explicitly structure work sequences, decision-making authority, and role division, and more \cite{schmidt1996coordination, hinckley1994survey, rico2008team}. However, these approaches have limitations, as they impose significant time, cognitive, and social costs on individuals by requiring the iterative organization and explanation of work outputs, as well as additional documentation and feedback coordination \cite{feldhoff2019bridging, lee2007boundary}.}

\re{In this context, GenAI presents significant potential to mitigate these challenges. Through translation between different modalities, such as text and image, GenAI can support participants in reducing friction in meaning interpretation and approaching mutual understanding while maintaining their preferred creative languages \cite{ko2022we, johanson2023if}. Moreover, unlike traditional boundary objects which tend to remain as static outputs that are costly to update \cite{langley2019boundary}, GenAI holds the potential to function as a fluid medium that is continuously modified and concretized across repeated interactions, as context accumulates over time \cite{wang2025social}. This can alleviate the formalization costs borne by humans and accelerate the process of meaning-making through interaction \cite{guridi2025fake}. Furthermore, GenAI provides an impersonal mediation process that complements face-to-face interactions prone to emotional exhaustion or hierarchical pressure, thereby efficiently managing the social costs of grounding \cite{claggett2025relational}. Despite this potential, empirical understanding of how GenAI can mediate collaborations among heterogeneous creators remains limited.}

\re{To examine the mediating potential of GenAI, this research selected the early stage of game development as the study context, where the difficulty of creative collaboration is maximized. This phase serves as a high-stakes testbed where the friction in building common ground between creators with distinct epistemic systems is most clearly revealed. In the early game development phase, narrative writers share a vision using narrative documents containing text and visual references as boundary objects, and visual designers translate these into visual language while repeatedly tuning for intentional alignment. This workflow implies high collaboration burdens in two aspects. First, epistemic asymmetry becomes pronounced as text-centric thinking and image-centric thinking intersect \cite{young2000creative}, generating substantial cognitive costs in interpreting boundary objects. Second, remote/asynchronous work conditions and frequent pivoting make real-time coordination difficult \cite{borg2019video, aleem2016critical}, imposing additional social and time costs on existing coordination mechanisms. These conditions provide a suitable environment to verify whether GenAI can operate as a mediator that complements or replaces collaboration coordination by accumulating context within asynchronous interactions.}

\re{Moreover, when GenAI is integrated into collaborations between heterogeneous creators and begins to take on the role of facilitating their work, it becomes necessary to explore a new set of concerns \cite{wei2025effects}. Existing research has primarily focused on existential concerns in the one-to-one creator-GenAI relationship—such as creator replacement \cite{mirowski2023co}, threats to creative identity \cite{kim2024unlocking, boucher2024resistance}, and loss of autonomy \cite{lee2022coauthor, gero2024writers}. Other studies have highlighted technical concerns with GenAI systems themselves, such as their unpredictable and biased outputs \cite{luo2024critical, mirowski2023co, newman2024want}, potential for stereotypes, and plagiarism \cite{ko2023large}. However, when GenAI operates between human collaborators, the focus of concern shifts to a relational and procedural dimension: how GenAI influences the relationships and processes among collaborators, and what concerns creators themselves raise about GenAI in this context. These new potential concerns remain underexplored.}

Therefore, we pose the following research questions:
\begin{itemize} 
\item RQ 1. What collaboration challenges do heterogeneous creators face, and what design opportunities do they envision for GenAI to better support their collaboration? 
\item RQ 2. When GenAI is integrated into their collaborative workflows, what potential concerns do creators perceive? 
\end{itemize}
To answer these questions, we conducted a multi-method study with narrative writers (N=10) and visual designers (N=10), including preliminary interviews and participatory design studies. Specifically, through co-design using designed stimuli—such as GenAI feature cards and creative prompting sessions—we helped participants concretely imagine and discuss the unfamiliar role of GenAI as a collaborative facilitator and mediator.

\re{We aim to contribute to CSCW and HCI through an empirical study on the role of GenAI in facilitating and mediating the collaborative processes of multiple creators. Our primary contributions are as follows: (1) We closely examine the collaboration between narrative writers and visual designers in the early game development stage to identify the fundamental challenges arising in heterogeneous creative teamwork across different modalities; (2) We propose design opportunities and guidelines for GenAI collaboration tools that facilitate and mediate creator relationships and mutual understanding throughout the entire collaborative process; (3) We identify potential concerns and key considerations that emerge when GenAI is integrated into collaborative workflows, offering practical insights for future tool design; and (4) We introduce an effective co-design methodology and creative-promting session that incorporates domain experts’ insights. Overall, these contributions collectively inform the design of GenAI tools, particularly in muti-creator collaboration contexts.}

\section{\MakeUppercase{Related Work}}
\subsection{Supporting End-to-End Creative Collaboration with GenAI}
Recent research on GenAI-based creativity support has increasingly expanded its focus from assisting individual creators to supporting collaborative processes. While early studies focused on aiding individual creative processes \cite{leandro2024geneva, anderson2024homogenization, girotra2023ideas, shen2024neural, oh2024lumimood, hou2024c2ideas, huang2024plantography} and designing workflow-optimized interactions \cite{choi2024creativeconnect, son2024genquery, sun2024generative, zhou2024understanding, moruzzi2024user, adobe2024generative, ko2023large}, recent work, leveraging GenAI's ability to streamline decision-making and consensus-building \cite{shin2022chatbots, hancock2020ai}, has seen emerging attempts to utilize it as a team member or a facilitator of collaboration \cite{shaer2024ai, gonzalez2024collaborative}. For instance, Houde empirically analyzed how the timing, method, and distribution of control for a GenAI teammate's interventions affect a group's idea generation process \cite{houde2025controlling}. Concurrently, Wang et al. showed that by analyzing accumulated patterns of individual interests, reactions, and participation, GenAI can identify common ground among participants and provide tailored materials to facilitate smoother collaboration and conversational flow \cite{wang2025social}.

\re{Despite this progress, how GenAI can mediate collaboration between creators with heterogeneous backgrounds remains a challenging task \cite{muller2019learning}. Participants with differing professional domains \cite{munster2014common, d2017comfy}, goal and perspectives \cite{billinghurst1999asymmetries} incur continuous communication costs to build common ground \cite{gardiner2020learning, feast2012professional,nguyen2020multidisciplinary}, a challenge that is further intensified under structural conditions such as remote, asynchronous environments or team hierarchies \cite{koschmann2003reconsidering, olson2000distance, gray2022confronting, grimm2024building}. Traditionally, the HCI and CSCW community has reduced collaborative friction by utilizing boundary objects, shared artifacts that bridge the epistemic gap between distinct creative domains, allowing collaborators to translate abstract concepts into a unified vision \cite{lee2007boundary, gonzalez2025achieving, saad1996shared}, or through coordination mechanisms that systematically manage the procedural flow and complex interdependencies among participants \cite{schmidt1996coordination, hinckley1994survey, rico2008team}.}

\re{Recent researchers are revisiting these classical CSCW methodologies by leveraging GenAI's capabilities in multimodal interpretation \cite{ko2022we, johanson2023if}, context augmentation \cite{wang2025social}, and consensus building \cite{shin2022chatbots, hancock2020ai}. First, GenAI is being utilized as a tool for generating boundary objects that assist in knowledge sharing by resolving asymmetries in expressive competence. For example, Ko et al. converted ambiguous ideas into concrete visual boundary objects by visualizing writers' text drafts for designers \cite{ko2022we}, while Ling et al. supported game designers with limited visualization skills in manifesting their ideas as illustrations \cite{johanson2023if}. However, these approaches were limited to unidirectional information transfer by focusing on augmenting individual expressive capabilities, and failed to sufficiently address the bidirectional consensus process of interacting around boundary objects to build common ground. Second, there are attempts to utilize GenAI as a coordination mechanism or real-time mediator to reduce friction in the collaborative process. Jose et al. demonstrated that the process of a facilitator or creator developing GenAI-generated ‘imperfects’ acts as a dynamic boundary object that promotes communication \cite{guridi2025fake}. Meanwhile, Claggett et al. proposed a Social Coordination model where an LLM identifies the state of social alignment during conflicting opinions and suggests appropriate phrasing \cite{claggett2025relational}.}

\re{However, these prior studies are limited by their focus on fragmentary phases of collaboration—primarily synchronous ideation sessions or real-time conversations. Yet, complex projects in real-world creative industries are predominantly conducted in remote and asynchronous manners due to spatio-temporal constraints \cite{morrison2020challenges, kleinsmann2008barriers}. Participants rarely meet in real-time, instead communicating intermittently by creating and delivering outputs in their own timeframes. In such asynchronous environments, where immediate feedback or social attunement is difficult, an approach that diachronically manages the shifting asymmetries across the entire process—from preparation to interaction to utilization—is essential. For instance, information asymmetry may serve as the primary challenge during the preparation phase, whereas power asymmetry might hinder communication during the interaction phase. Yet, a systematic exploration of these dynamic asymmetries and an integrated approach to support the full collaborative lifecycle are still underexplored.}

\re{Therefore, this study aims to fill this gap by examining the collaboration process between narrative writers and visual designers in the early stages of game development. In this phase, there exists a fundamental asymmetry of conflicting languages and goals \cite{johanson2023if}, with writers prioritizing narrative consistency while designers prioritize visual coherence \cite{young2000creative}. This setting provides an ideal research environment where spatio-temporal constraints, information imbalances, and organizational power asymmetries complexly intersect. The tight deadlines characteristic of the game industry serve as factors that further exacerbate these conflicts \cite{borg2019video, aleem2016critical}. We explore how GenAI tools can function as a new type of Boundary Object and Mediator that facilitates meaning-making and coordination in interdisciplinary collaboration.}

\subsection{Perception and Concerns About Collaborating with GenAI in Creative Process}
The discourse surrounding GenAI in creative fields has been shaped by a fundamental tension between its potential as a tool and the threat it poses to human creators \cite{kim2024unlocking, esmaeili2024strategize}. Creators' anxieties about GenAI have given rise to various concerns that act as a key barrier to technology adoption \cite{mirowski2023co}, and the HCI community has focused on identifying the substance of these concerns.

The concerns identified in prior research can be broadly divided into two categories. The first involves threats to the creator's agency and professional identity. This includes the fear of creative roles being replaced \cite{mirowski2023co}, the loss of autonomy in the creative process \cite{lee2022coauthor, gero2024writers}, and worries about the loss of a ‘human touch' and the dilution of unique identity \cite{kim2024unlocking, boucher2024resistance}. The second category comprises risks stemming from the opacity and unreliability of the AI system itself. The unpredictable and biased outputs of AI \cite{luo2024critical, mirowski2023co, newman2024want}, the reproduction of stereotypes and potential for plagiarism \cite{ko2023large}, and its inability to understand social nuances \cite{gero2023social} are key technical and ethical barriers that make it difficult for creators to accept AI as a trusted partner.

While prior work has extensively identified the existential and technical concerns arising from the personal creator-AI dyad, providing crucial insights for designing GenAI systems for individual creators, the emerging context where GenAI's role expands to that of a facilitator and mediator for collaboration between heterogeneous experts remains largely underexplored. This shift in role may require our inquiry to expand, allowing us to revisit established concerns in this new context and to discover new ones that were previously unaddressed \cite{johnson2025exploring, bangerl2025creaitive, doshi2024generative}. Specifically, how creators perceive this new role—and what relational and procedural concerns they experience within their collaborative practices—remains a significant gap in the literature. This study, therefore, aims to fill this gap by empirically investigating the specific nature of concerns that arise when GenAI functions as a facilitator and mediator in a heterogeneous creative environment.

\subsection{Co-design Method for User-Centered GenAI}
The Co-design method is one of the most effective approaches for designing digital technologies from the end-user's perspective \cite{harrington2019deconstructing, zhang2022algorithmic, pierre2021getting, kim2024unlocking}. By involving users from the initial design stages, this approach helps in revealing hidden problems and discovering user-centered solutions. Solutions designed based on users' domain insights help users more easily accept and adapt to new technologies \cite{peek2016takes, scandurra2013participatory}.

With the integration of AI into systems, the co-design method has faced new challenges, such as AI's unpredictability \cite{liao2023designerly, yang2020re} and the difficulty of aligning AI technology with human needs \cite{moore2023failurenotes}. In response, the HCI community has focused on helping users better understand AI's capabilities and ideate ways to integrate it into their workflows \cite{AccentureTheDock2020, gorkovenko2020exploring, jin2021design, kun2018design, malsattar2019designing}. Recently, one study built upon this research to design co-design sessions that prompted domain experts to devise AI-driven ideas for solving problems in their creative processes using stimuli \cite{kim2024unlocking}. This study utilized four types of stimulus materials to help participants successfully generate appropriate AI ideas.

Inspired by this research, we structured our co-design activities, but our methodology diverges in two significant ways. First, many widely used stimuli in previous work were developed with predictive or analytical AI in mind \cite{yildirim2023creating, zhou2020ml, amershi2019guidelines, google2019people}, and are thus limited in addressing the unique expressive and interactive modalities of today’s generative AI. Second, as previously discussed, existing approaches primarily focus on AI that assists individual task performance, meaning there is scarcely an established process for co-designing a generative AI that interacts and mediates within a collaborative context.

This study devised and executed a participatory design process to guide participants in understanding the characteristics and potential of GenAI and applying them to collaborative problems. We provided ‘GenAI Capability Cards,' a set of stimuli updated for the generative paradigm. We also designed ‘Creative Prompting Sessions' to allow participants to explore the potential of GenAI across their entire workflow by experiencing interaction points with state-of-the-art GenAI tools.

The stimulus materials we proposed were as follows: (i) GenAI Capability Cards: Cards to help participants easily understand GenAI features, (ii) Interventions for GenAI Concept Formation, (iii) Creative Prompting Sessions: Sessions designed for participants to experience and understand interaction points by using the latest GenAI tools in a complete workflow, and (iv) A template for organizing AI feature and task ideas: A series of questions to help creators articulate both the positive effects and the concerns associated with their ideas \cite{O_Loideain2019}.
Through this structured co-design process, we not only enabled participants to propose user-centered ideas but also effectively captured the nuanced concerns and collaboration-specific design requirements that arise when GenAI intervenes as a mediator in creative collaboration.

\section{\MakeUppercase{Method}}
\subsection{Positionality Statement}
Our author team consists of researchers with academic training and research experience in design, human–AI interaction, and responsible AI, as well as diverse expertise in AI design and development practices. Two authors have conducted extensive research on human–AI interaction design and responsible AI, drawing on practical experience with user concerns and experiences around the creative use of AI. Another author has research experience in building and designing AI systems as creativity-support tools. These backgrounds and experiences inform our positionality and shape the subjectivities inherent in our research questions, study design, participant recruitment, and data analysis.

\subsection{Overview of the Multi-Method Study}
We first conducted preliminary interviews with four narrative writers and four visual designers to understand collaboration in early game development. These sessions revealed differences in their creative processes and provided guidance for structuring our study. Building on these insights, we designed a creative-prompting session that integrated GenAI into creators’ workflows. We then ran a participatory design study with 14 creators to examine major collaboration barriers and AI’s potential roles. The findings highlight how GenAI can support creativity and efficiency, enhance communication across roles, and surface concerns around its use in creative workflows.

\subsection{Participants}
\begin{table}[t]
    \centering
    \captionsetup{justification=centering}
    \renewcommand{\arraystretch}{1.5}
    \begin{tabularx}{0.9\textwidth}{@{\hskip 0.5cm} c >{\centering\arraybackslash}p{2cm} >{\centering\arraybackslash}p{2cm} >{\centering\arraybackslash}p{1.5cm} >{\centering\arraybackslash}p{1.5cm} >{\centering\arraybackslash}p{3cm} @{\hskip -0.5cm}} 
        \hline
        \specialrule{1pt}{0em}{0em}
        \textbf{\makecell{Num \\}} 
        & \textbf{\makecell{Preliminary \\ Interview}} 
        & \textbf{\makecell{Design \\ Study}} 
        & \textbf{Gender} & \textbf{Age} & \textbf{\makecell{Work \\ Experience}} \\
        \hline
        PW1 & \checkmark &  & Male & 32 & 4 years \\
        \hline
        PW2 & \checkmark &  & Male & 46 & 14 years \\
        \hline
        PW3 & \checkmark &  & Male & 42 & 4 years \\
        \hline
        PD1 & \checkmark &  & Female & 29 & 6 years \\
        \hline
        PD2 & \checkmark &  & Female & 29 & 8 years \\
        \hline
        PD3 & \checkmark &  & Male & 33 & 10 years \\
        \hline
        \specialrule{1pt}{0em}{0em}
        W1 &  & \checkmark & Male & 43 & 15 years \\
        \hline
        W2 &  & \checkmark & Male & 35 & 5 years \\
        \hline
        W3 & \checkmark & \checkmark & Male & 54 & 10 years \\
        \hline
        W4 & & \checkmark & Female & 45 & 3 years \\
        \hline
        W5 & & \checkmark & Male & 44 & 11 years \\
        \hline
        W6 & & \checkmark & Female & 32 & 1 year 10 months \\
        \hline
        W7 & & \checkmark & Female & 36 & 9 years \\
        \hline
        D1 & & \checkmark & Male & 35 & 4 years \\
        \hline
        D2 & & \checkmark & Female & 26 & 1 year 5 months \\
        \hline
        D3 & & \checkmark & Male & 29 & 1 year \\
        \hline
        D4 & \checkmark & \checkmark & Male & 43 & 17 years \\
        \hline
        D5 & & \checkmark & Female & 29 & 5 years 10 months \\
        \hline
        D6 & & \checkmark & Female & 30 & 2 years \\
        \hline
        D7 & & \checkmark & Male & 26 & 2 years \\
        \hline
        \specialrule{1pt}{0em}{0em}
    \end{tabularx}

    \caption{Participant information summary, detailing involvement in the preliminary interviews and the participatory design study, along with demographic data. Participants are labeled as follows: PW = Preliminary Writer, PD = Preliminary Designer, W = Writer (participatory design), and D = Designer (participatory design)}
\end{table}

We aimed to recruit creators who have substantial experience in narrative production or concept art in the game industry based on the following criteria: (i) individuals aged 19 or older, (ii) narrative writers or visual designers with at least one year of experience in the game industry, (iii) individuals who have participated in at least one game development project, and (iv) for narrative writers, at least three years of writing experience, and for visual designers, at least three years of design experience.

\re{The study recruited participants based in South Korea.} We employed a purposive sampling strategy supplemented with snowball sampling. Purposive sampling was necessary because our study required participants with professional experience in narrative writing or visual design for games, ensuring that participants could meaningfully reflect on collaboration practices and evaluate the role of GenAI in their workflows. Snowball sampling was additionally used to access professional networks that are less visible through public postings but highly relevant for this niche creative domain.

The recruitment of participants was conducted through four main channels: (1) social media platforms like LinkedIn and Instagram, (2) online community platforms of various universities, (3) outsourcing platforms for writers/designers like Kmong, and (4) snowball sampling. These four channels were chosen to capture different segments of the creator community: industry professionals active on LinkedIn/Instagram, early-career creators in university networks, freelancers and independent contractors on Kmong, and additional experienced professionals accessed via referrals. This combination aligned with our goal of recruiting a diverse range of creators engaged in early game development. To ensure a diverse range of experiences among game narrative writers and visual designers, we included a pre-screening survey link in recruitment posts and invitation emails. \re{The survey asked participants to report their professional work experience in the game industry (e.g., 5 years) as narrative writers or visual designers, along with the AI tools they use in their professional practice.}

A total of 20 participants were recruited. 8 participants took part in the preliminary study, and 14 participated in the participatory design study, all of whom met the recruitment criteria. Two participants were involved in both the preliminary study and the participatory design study. All participants were self-reported, but their experience was verified through knowledge outsourcing platforms, LinkedIn and Kmong, and reference checks. Recruitment was concluded once we reached the target sample size and achieved a balanced distribution of participant roles, while also considering practical constraints such as budget and scheduling.

Among the 8 participants in the preliminary study, four were narrative writers, and four were visual designers. The average age of the narrative writers and visual designers was 43.5 years (SD = 9.15) and 33.5 years (SD = 6.6), respectively. Among the fourteen participants in the participatory design study, seven were narrative writers and seven were visual designers. The average age of narrative writers and visual designers was 41.28 years (SD = 7.52) and 31.14 years (SD = 6.04). \re{Participants were asked about their prior experience using GenAI tools in professional work contexts. Among the participating narrative writers, five reported prior experience using GenAI tools, while two had no such experience. Those with experience primarily used text-to-text tools such as ChatGPT. Among the visual designers, six reported prior experience using GenAI tools, whereas one had not. Designers with GenAI experience reported using a wider range of tools and modalities, including text-to-text tools (e.g., ChatGPT) as well as text-to-image and image-to-image tools (e.g., Midjourney, NovelAI, and Stable Diffusion).}

Detailed information is referred to in Table 1, which provides an overview of participants' fields and roles. Participants are identified by [role abbreviation][role number]: writers (W\#), and designers (D\#).

Participants in the preliminary interview received a coffee coupon (worth \$3.85) as compensation, while participants in the participatory design study received 100,000 KRW (approximately \$73 USD). All study ensured confidentiality and the protection of personal identifying information for all participants, and it was explained that all responses would be anonymized. Additionally, participants were informed that they could skip any uncomfortable questions or leave the activities at any time, for any reason.

This study was conducted in accordance with established research ethics protocols. The study protocol was reviewed and approved by the Institutional Review Board (IRB) prior to data collection. All participants were informed of the study’s purpose, procedures, and their rights before providing written consent. Specifically, participants were told that (i) their participation was voluntary, (ii) they could skip any question or withdraw from the study at any point without penalty, and (iii) all collected data would be anonymized and stored securely. To protect participants’ privacy, video recordings were made during online sessions, and all identifying information was removed during transcription and analysis. Throughout the study, we emphasized respect for participant well-being and confidentiality, aligning with ethical standards for human–computer interaction research.

\subsection{Preliminary Interviews}
\re{We conducted preliminary interviews with four narrative writers and four visual designers. These interviews were conducted as individual, one-on-one sessions. The purpose of these interviews was to gain a deep understanding of the individual workflows and collaborative processes of writers and designers, with the goal of designing a participatory design study process that reflects their creative workflows. The interviews lasted one hour, and the questions were focused on their personal and collaborative production processes and their experiences with GenAI tools. All interviews were recorded via Zoom and transcribed using Clova Note6 and DaGlo, with cited quotes translated into English.} 

Using qualitative thematic analysis \cite{braun2019reflecting}, three researchers independently coded the transcripts, resolving discrepancies through team discussions for consistency. The codes were then iteratively organized into themes that informed the production process and insights for participatory design study. These interviews provided critical insights into creators’ production processes (Section 4.1) and guided the structure and planning of the participatory design study to align with participants’ preferences and work styles. Key insights for shaping the participatory design study are summarized as follows:

First, we found that creators with limited experience of using AI often struggled to envision how AI could be integrated into their collaborative processes. To address this, we incorporated a ‘creative-prompting’ activity into the participatory design study, allowing participants to explore the latest GenAI tools within the context of their own work.

To further support participants with limited experience using GenAI, we developed tutorial materials after the preliminary interviews, demonstrating step by step how to integrate ChatGPT-4o and MidJourney Alpha into their workflows. These tutorials were specifically tailored based on the insights gained from the preliminary interviews, ensuring that participants could easily follow along and replicate the steps. After the interview, the tutorials were provided in various formats, including text, comic, and video to accommodate participants’ preferred learning styles and to offer flexibility in their engagement. Consistently, participants found the video format to be the most accessible and effective, aligning with research suggesting that video-based learning materials are more effective than text-based materials for information delivery and comprehension \cite{lin2023meta}. Based on these findings, we decided to offer instructional tutorials on ChatGPT-4o, MidJourney Alpha, and prompt creation in video format.

Furthermore, the study revealed that writers, especially those with prior experience using MidJourney, generally prefer text-based communication over GUI-based tools. Based on this insight, we chose to focus solely on ChatGPT for writers in the main study. 

\re{Importantly, the preliminary interviews also revealed that early-stage collaboration between narrative writers and visual designers is often remote and asynchronous rather than conducted through real-time, co-located collaboration. Participants described working independently on their respective artifacts (e.g., narrative documents or concept images) and exchanging these outputs through remote channels such as shared documents, messaging tools, or email, rather than engaging in continuous joint co-creation sessions. Several participants highlighted challenges arising from this collaboration structure, including temporal delays, limited opportunities for real-time clarification, and difficulties in conveying creative intent across roles.}

\re{Based on these insights, we deliberately designed the creative-prompting sessions as individual sessions to reflect creators’ existing collaborative practices and to examine how GenAI could mediate collaboration across separated roles and asynchronous workflows, rather than assuming synchronous, paired co-creation that does not reflect participants’ everyday work contexts.}

\begin{figure}[t]
    \centering
    \includegraphics[width=\textwidth]{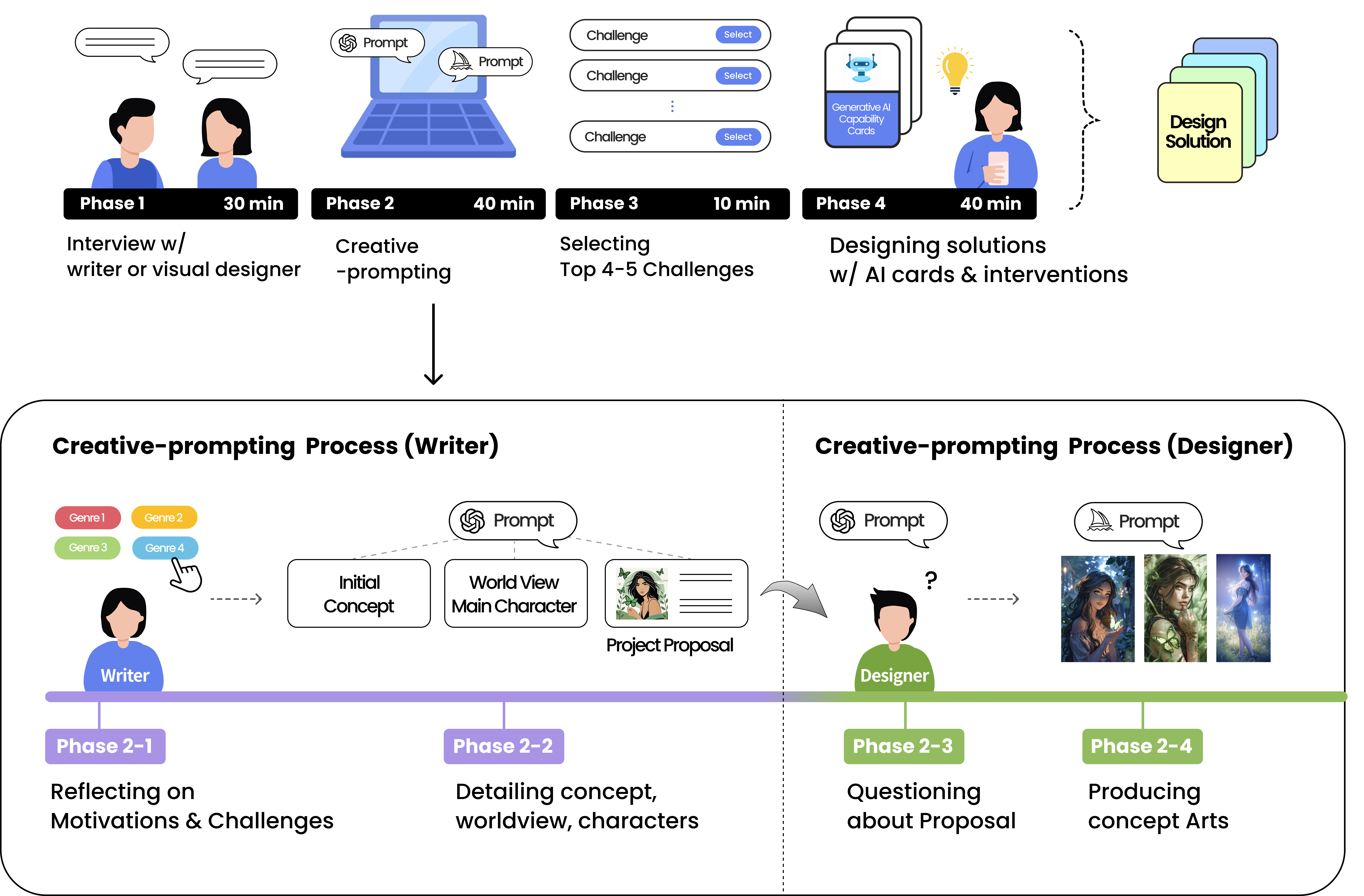}
    \caption{\re{The image at the top represents an overview of our participatory design study protocol, while the image below specifically illustrates the creative prompting process.}}
    \label{fig:workflow}
\end{figure}

\subsection{Participatory Design Study Protocol}
This study aims to propose design suggestions for how GenAI can improve the production and collaboration processes of narrative writers and visual designers. To achieve this, we first conducted semi-structured interviews to investigate the communication challenges arose between the narrative writers and visual designers, stemming from the unique characteristics of game production and their differing areas of expertise. Based on the insights gathered from these interviews, we then conducted a participatory design study to address the identified challenges. The study combined semi-structured interviews and a participatory design study (see Figure 2, top), with each session lasting approximately two hours.

In light of the COVID-19 pandemic, and some creators preferred not to participate in face-to-face studies, all interviews and activities were conducted remotely via Zoom. Only audio recordings (no video) were taken, which were later transcribed for analysis. Each interview and activity involved one participant, with one researcher acting as the facilitator and another researcher assisting.

During the interviews, participants were asked to verbally explain the challenges they faced in their individual and collaborative work. Prior to attending the participatory design study, participants were provided with tutorial materials introducing the use of GenAI tools. The participatory design study then focused on collaborating with participants to design solutions to these problems. The entire process of the participatory design study is illustrated in Figure 2. One researcher specialized in AI research, while the other specialized in AI user experience, both having strong backgrounds in the field of HCI. The structure of the participatory design study was inspired by previous research \cite{kim2024unlocking, zhang2022algorithmic} focused on collaborative workshops to generate new solutions.

\textit{\textbf{Introduction and Icebreaking (5 min):}}
We introduced the purpose of the study, the goals of the activities, and the procedures. To build rapport and mutual understanding, participants were asked to introduce themselves by sharing their name or nickname, area of expertise, and game content creation experience.

\textbf{\textit{Phase 1: Defining Challenges (25 min):}}
Through focused interviews, we defined the challenges participants face. The participants verbally explained the difficulties they encounter in collaborative work.

\textit{\textbf{Phase 2: Completing Work Through AI Tool Prompting (40 min):}}
\re{Based on the preliminary interview, we designed a ‘creative-prompting’ session to help participants imagine how AI could be integrated into their workflow (See Figure 2, bottom). Before the session began, researchers provided detailed explanations of the AI tools and introduced prompting as a general interaction practice for using GenAI systems. Participants then created their own prompts, with researchers actively encouraging questions and open communication. These efforts were intended to minimize confusion during the prompt creation phase.}

\re{During the session, first, the narrative writer closely collaborated with ChatGPT-4o to create a narrative, writing a game narrative document for the character and generating a simple main character image. Then, to simulate the process in which an designer would ask questions via email or in meetings about the game narrative document, the writer shared the conversation log with ChatGPT-4o to the designer. \re{The designer could then ask follow-up questions to ChatGPT-4o within the conversation log shared by the writer to resolve any uncertainties.} Afterward, the designer used MidJourney alpha to visually develop the main character based on their understanding. This process provided participants with practical experience of how AI could be applied not only in individual work but also in collaborative settings. Unlike previous studies \cite{sarkar2023participatory} that utilized participatory prompting—a method in which researchers craft prompts by interacting with AI, present the outputs to users for feedback, and iteratively refine the prompts based on that feedback—our study takes a novel approach. We empower creators to independently craft their own prompts and directly integrate them into their creative workflows. Example prompts created by participants are provided in Appendix~\ref{appendix:exampleprompts}.}

\textit{\textbf{Phase 3: Evaluating Challenges and Sharing Experiences (10 min):}}
\re{Two researchers participated in the activities—one as the facilitator and the other as the note-taker, whose role was to document the challenges discussed by participants. Based on the challenges identified during Phases 1 and 2 within each session, we constructed a survey that included an inclusive list of collaboration challenges raised in that session. Accordingly, the survey content varied across sessions rather than relying on a fixed or predefined challenge set. Through a self-evaluation questionnaire, participants selected the most important 4-5 challenges they personally faced. From these participant-selected challenges, the researchers then chose two key challenges to frame collaborative reflection activities, which were presented as missions. Participants shared their experiences and reflections on why these challenges were significant, generating insights and inspiration for subsequent analysis.}

\textit{\textbf{Phase 4: Designing Solutions to Address the Challenges (40 min):}}
Participants were provided with GenAI capability cards, an introduction to the latest GenAI tools, and an idea summary template to assist them in devising solutions (See Appendix \ref{appendix:capabilitycards}). The GenAI capability cards, inspired by prior research \cite{yildirim2023creating}, were designed to help participants understand GenAI technology and support brainstorming. These cards were created based on existing studies \cite{lyu2024preliminary,shi2023hci,weisz2023toward}, making it easier for participants with limited GenAI background knowledge to imagine its functions through everyday language and visual materials. During the experiment, the GenAI cards were presented in the form of PowerPoint slides. Each function and example of GenAI was explained individually, and participants were encouraged to either build on the cards provided or freely ideate about additional GenAI capabilities not included in the cards. Participants used these slides as a reference while brainstorming ideas.

Additionally, examples of the latest GenAI tools were introduced to encourage participants to explore new possibilities, even with GenAI features they had not previously experienced. Finally, participants utilized the Insight-Action-Outcome idea summary template to reflect on the insights gained from challenging situations, define the actions AI could take to address these challenges, and organize specific solutions in a structured manner. They were also asked to consider and discuss potential side effects of their proposed solutions.

Through these systematic steps, participants were able to understand GenAI technologies and develop design suggestions.

\subsection{Data Analysis}
We recorded our participatory design study sessions via Zoom and transcribed the audio using Clova Note6 and DaGlo. All cited quotes were translated into English. We analyzed the data using a reflexive thematic analysis approach \cite{braun2019reflecting}. In the initial open coding stage, three researchers carefully read the transcripts, took notes, and coded the data. Each researcher independently generated initial codes from participants’ accounts. This process resulted in 538 codes. Each transcript was reviewed by at least two researchers, and the team held regular meetings to compare insights and discuss emerging ideas.  

Through an iterative process, we clustered codes into higher-level themes related to collaboration challenges, clarifying the difficulties and needs in collaborative work. We also analyzed design solutions, categorizing them into high-level themes based on their purposes and the specific challenges they addressed. For instance, one initial code noted: “D4 suggested that a conversational AI that understands and retrieves information from documents uploaded by writers would be highly helpful, reducing the need to ask writers directly and easing psychological burdens.” Such codes were grouped into broader, overarching themes—for example, “Facilitating Information Accessibility and Mediation through GenAI.” We further identified recurring patterns reflecting concerns about integrating GenAI into collaborative work.

\re{The research team held regular collaborative sessions whenever new codes emerged, focusing on discussing their meaning, scope, and relationships to other codes. When particular excerpts allowed for multiple possible interpretations, the researchers returned to the original transcripts to re-examine how participants framed their experiences in context. Interpretive ambiguities or disagreements were addressed through these ongoing discussions by refining code definitions or adjusting boundaries between codes. The rationale for these analytic decisions was documented in analytic memos, through which themes were iteratively refined and interpretive differences were reconciled. Following the principles of reflexive thematic analysis, we did not calculate inter-rater reliability, as collaborative interpretation and repeated team discussions were central to generating codes and themes \cite{braun2019reflecting, mcdonald2019reliability}.}

\re{Finally, this process generated 65 first-level themes, which we clustered into 22 second-level themes and then synthesized into 3 third-level themes. These three overarching themes capture: (i) challenges in creative work, (ii) corresponding design solutions, and (iii) the creators’ perceptions and concerns of using GenAI in creative collaboration. We present these themes in detail below and summarize them in Table~\ref{tab:themes_summary_genai}.}

\section{\MakeUppercase{Findings}}
We divided the findings into three main sections:  
(1) the creative process and collaboration between narrative writers and visual designers, (2) the challenges within these collaborative processes and design suggestions for GenAI tools to address them, and (3) creators’ concerns about GenAI’s intervention in creative work and collaboration. We present the challenges and design suggestions (Section~\ref{sec:cands}) in alignment with the production process so that the themes follow the actual sequence of creators’ workflows. We then explain broader concerns about integrating GenAI into creative collaboration across all stages of the workflow (Section~\ref{sec:concerns}).

\subsection{Production Process}
To create GenAI tools that support the creative production and collaborative production processes between narrative writers and visual designers, it is essential to first explore the production process in the early stages of game development. Through our analysis of the preliminary interview, we organized the early stages into three key levels:

\subsubsection{Narrative Construction Phase}
In the initial phase, the narrative writer generates foundational story ideas and preliminary settings (e.g., game universe, characters, and backgrounds) aligned with the game's genre and core concepts. The writer refines the narrative direction and outlines key elements, laying the groundwork to enable clear communication and effective collaboration with visual designers in subsequent stages.

\subsubsection{Game Narrative Document Communication Phase}
In this phase, the narrative writer translates the developed story and its key elements into a detailed Game Narrative Document. This document includes concise descriptions of abstract visual elements and is supplemented with relevant reference images sourced online. Its purpose is to ensure the visual designer clearly understands the intended narrative, mood, characters, and visual style, minimizing the risk of misinterpretation and guiding visualization.

\subsubsection{Visual Concept Art Creation Phase}
In the final collaborative stage, the visual designer creates visual concept drafts based on the Game Narrative Document. The designer clarifies any ambiguous aspects through direct communication with the narrative writer, while independently gathering background knowledge and visual references to maintain historical and thematic accuracy. Typically, the designer prepares several rough drafts, iteratively refining them through feedback from the narrative writer. Once a final draft is selected, the designer meticulously details the artwork by defining composition, outlines, colors, lighting, shading, and intricate visual textures.

\subsection{Challenges and Design Opportunities in
Creative Workflow and Collaboration}
\label{sec:cands}
\begin{figure}[t]
\centering
\includegraphics[width=\textwidth]{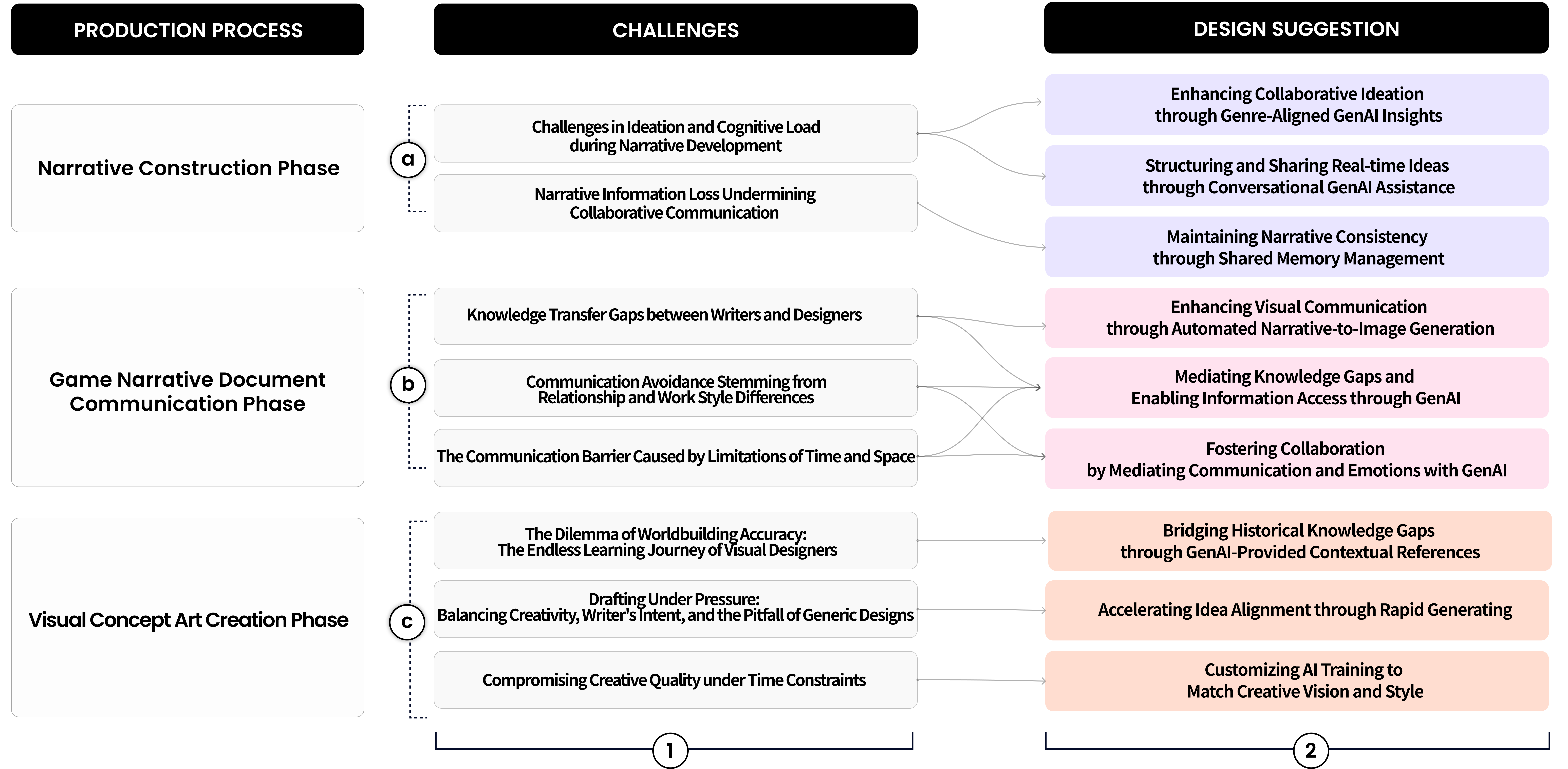}
\caption{\re{
Thematic analysis map connecting identified challenges and design suggestions from participatory design sessions.
(a) Challenges in narrative writing.
(b) Challenges in game narrative document communication.
(c) Challenges in visual concept art creation.
\textit{Left (1)} shows the synthesized and consolidated challenges identified and prioritized by participants across sessions. \textit{Right (2)} presents design suggestions proposed during the participatory design sessions. The two purple design suggestions specifically address narrative writing challenges (a).
The three pink design suggestions specifically target communication challenges in game narrative documents (b). The three orange design suggestions focus explicitly on visual concept art creation challenges (c).
}}
\label{fig:your_label}
\end{figure}

In this section, we outline the challenges narrative writers and visual designers encounter during each stage of their production process. For each stage, we provide tailored design suggestions for future GenAI collaborative tools to address these challenges. Our following insights not only present these challenges but also demonstrate how each design opportunity directly addresses specific challenges and collaboration asymmetries. This mapping between challenges and design opportunities is illustrated in detail in Figure \ref{fig:your_label}.

\subsubsection{Challenges in Narrative Construction Phase}{\textit{ }}

\re{\textit{\textbf{Challenges in Ideation and Cognitive Load during Narrative Development.}}
In the preparation phase for collaborative ideation, writers face intensive ideation demands as they attempt to align story concepts with gameplay mechanics and player experience. While many ideas emerge from abstract and personal sources of inspiration, these ideas often remain unstable and insufficiently grounded in shared genre conventions, target audiences, or gameplay constraints. As a result, early ideation frequently fails to produce concrete, shareable reference points that collaborators can easily align around. Moreover, writers must repeatedly generate, discard, and revise ideas while also bearing the burden of externally structuring and documenting them for collaboration. Spontaneous ideas are often lost or delayed before they can be formalized and shared, increasing cognitive load and creating friction at the very entry point of collaboration.}

\re{\textit{\textbf{Narrative Information Loss Undermining Collaborative Communication.}}
Information asymmetry often originates from narrative writers’ difficulty in consistently remembering and managing extensive narrative details accumulated over time. As a result, narrative documents may contain incomplete or inaccurate information from the outset, leaving designers to work with gaps or inconsistencies. Such narrative information loss significantly undermines consistency and creates confusion in later collaboration stages. W5 explained, \textit{"There are so many different settings, and it's easy to forget or struggle to remember all the details. So, I have to keep separate records in Excel sheets for characters, backgrounds, and other settings, checking them repeatedly."} Designers often need to request additional clarification from writers due to incomplete or ambiguous details. Visualizations produced under such conditions risk introducing narrative inconsistencies that undermine world-building quality, negatively affecting the overall quality of the final product.}

\subsubsection{Design Suggestions}{\textit{}}

\begin{itemize}
\item \textit{\textbf{Enhancing Collaborative Ideation through Genre-Aligned GenAI Insights.}}
\re{Participants noted that collaborative ideation often breaks down when early ideas are not clearly aligned with the game’s genre, target audience, and gameplay mechanics. To address this, they envisioned GenAI as a mediator that could provide genre- and gameplay-aligned ideas, helping narrative writers and visual designers align around a shared starting point. For example, W4 stated, \textit{“It would be great to have an AI that can develop detailed ideas based on the unique story rules (plots) and character attributes specific to genres such as fantasy, sci-fi, or martial arts.”} Similarly, W3 discussed first-person shooter (FPS) games, suggesting that GenAI could propose combat scenarios set in environments with natural cover, such as ruins or dense forests, to support strategic gameplay, while deliberately avoiding locations with limited tactical depth, such as open plains.}

\item \textit{\textbf{Structuring and Sharing Real-time Ideas through Conversational GenAI Assistance.}}
\re{Participants described a breakdown at the entry point of collaboration, where spontaneous ideas are often lost or delayed because they must be individually structured and documented before being shared. To intervene in this breakdown, they envisioned conversational GenAI that transforms fleeting ideas into minimally structured drafts that can be immediately shared with collaborators. As W6 explained, \textit{“If I could exchange ideas with an AI like ChatGPT and have it generate a structured draft, I could quickly share concept ideas with visual designers without spending additional time.”} By lowering the documentation barrier and enabling real-time sharing, participants expected such GenAI support to facilitate more flexible and timely alignment during the production process.}

\item \textit{\textbf{Maintaining Narrative Consistency through Shared Memory Management.}}
Participants valued GenAI's potential as a collaborative memory platform, capable of systematically storing and managing comprehensive narrative details and story decisions. W2 stated, \textit{“Honestly, it's very challenging for an individual to find desired information solely through search. You need to remember all the works you've previously encountered and accurately recall each work's settings and details. If AI could assist with memory and information management, you could instantly access the required information without having to revisit each work repeatedly.”} Participants hoped that even when narrative writers forget or inadvertently omit specific story details or settings, causing incomplete information transfer to designers, the GenAI system would systematically preserve extensive narrative data, making it easily accessible to all team members. Additionally, they envisioned GenAI automatically identifying contradictions in settings or inconsistencies in character information and promptly alerting collaborators to maintain narrative coherence. Such a GenAI-based shared memory system would significantly reduce repetitive clarification requests during collaboration, minimize errors from iterative revisions, and uphold overall narrative consistency, effectively enhancing the completeness and quality of creative outputs.
\end{itemize}

\subsubsection{Challenges in Narrative Document Communication and Collaborative Interaction} {\textit{}}

{\textit{\textbf{Knowledge Transfer Gaps between Writers and Designers.}}}
In the game narrative document creation and communication phase, there is the asymmetry of knowledge and expertise between narrative writers and visual designers. Therefore, the game narrative document should be crafted to help visual designers understand the core elements of the narrative from multiple perspectives, enabling them to better interpret and visualize the narrative. D6 noted, \textit{“Not only the visual elements but also comprehensive information such as personality and backstory must be provided to create a lifelike character design.”} However, narrative writers, being less familiar with design, often fail to convey the necessary details for visualization effectively. As a result, as seen in D4's case, missing information in the game narrative document is often filled in by the visual designer’s imagination. This leads to situations where, as W3 expressed dissatisfaction, \textit{“The drafts that designers bring always differ from the narrative writer’s original intention.”} 

To address this issue, narrative writers often search for reference images and attach them to the game narrative document. However, finding images that clearly convey their intentions is nearly impossible, often requiring compromises. D2, D3, and D6 argued that the reference images provided by narrative writers are often insufficient to fully capture the tone and atmosphere of the overall concept. W5 expressed frustration, saying, \textit{“If I could sketch, I’d be able to communicate my intent more effectively, but lacking that ability, I always feel limited in conveying my ideas through stock images.”}

{\textit{\textbf{Communication Avoidance Stemming from Relationship and Work Style Differences.}}}
\re{Collaboration was frequently affected by asymmetries in relationships and work styles, including hierarchical dynamics and differences in individual ways of working, which complicated communication across roles. Participants W6 and W7 pointed out that conflicts among team leaders or hierarchical requests for revisions cause issues. Delays in communication due to strained team relationships can stall work progress, while top-down directives from decision-makers unfamiliar with the relevant domain often blur work contexts, resulting in outputs that practitioners find difficult to empathize with. Participants W2, D4, and D7 further noted that concerns about encroaching on others’ expertise often lead to deliberate avoidance of questions or interactions, preventing timely repair of misunderstandings. Similarly, W5 emphasized that indirect communication mediated by team leaders (e.g., art directors), rather than direct exchanges between designers and writers, increases coordination overhead and obscures shared understanding.}

{\textit{\textbf{The Communication Barrier Caused by Limitations of Time and Space.}}}
\re{Collaboration was further challenged by asymmetries in the communication environment—with email and online chat often being the only channels outside of meetings—and by asymmetries in working conditions, as creators faced busy schedules with little overlap in their available time. These mismatches frequently delayed communication and timely feedback. As deadlines approached, participants reported hesitating to initiate additional communication to avoid increasing the workload of other teams, often proceeding with incomplete alignment. As a result, visual designers sometimes worked on drafts without fully understanding the narrative context (D2), while writers reported accepting drafts that differed from their original intent as long as they did not deviate significantly (W5).}

There are also instances where narrative writers do not have the opportunity to explain the game narrative document directly to visual designers. Email and online chats are the only communication methods outside of meetings, but due to busy schedules, responses to detailed questions or feedback are often delayed, resulting in production delays or late-stage revisions. To address this issue, D1 expressed the need for an environment where they can freely ask questions and communicate without the constraints of time and space, fostering a deeper mutual understanding of each other’s work. 

\subsubsection{Design Suggestions}{\textit{ }}

\begin{itemize}
\item{\textit{\textbf{Enhancing Visual Communication through Automated Narrative-to-Image Generation.}}} Participants envisioned GenAI collaboration tools that could support narrative writers in creating narrative documents by easily generating reference images from text-based materials, enabling more effective communication with designers. For instance, W1 expressed a desire for GenAI to automatically identify sections of the game narrative document that require visualization and generate relevant images. W3 added, \textit{“When searching for images or doing research, I often use Pinterest, but it’s difficult to find something that exactly matches the image in my mind. If GenAI could generate visuals based on the narrative document, I think it could produce results closer to what I’ve imagined. This would make communication with designers much smoother.”} In response, W2 and W6 suggested a system that combines the latest image generation tools with interactive agents. This system would enable GenAI to engage in real-time conversations based on the game narrative document, identifying sections where additional information is needed for visualization and generating corresponding images. It would also allow writers, even those unfamiliar with production tools, to refine these GenAI-generated visuals based on their textual input.

\item {\textit{\textbf{Mediating Knowledge Gaps and Enabling Information Access through GenAI.}}}
\re{Participants described GenAI as a mediator that could support collaborative communication by reducing knowledge gaps and communication delays. In particular, visual designers wanted GenAI to directly access narrative materials and provide task-relevant information in real time. D4 emphasized, \textit{“Maintaining the continuity of creative work is crucial. If writers could upload their documents to AI, a conversational interface enabling easy information retrieval through text-based questions would be incredibly helpful. This way, we wouldn’t need to wait to ask writers directly, reducing the psychological burden. If AI could take on that mediating role, it would be extremely valuable.”}}

\item{\textit{\textbf{Fostering Collaboration by Mediating Communication and Emotions with GenAI.}}}
Participants envisioned GenAI as playing a critical role in enhancing communication dynamics, resolving interpersonal conflicts, and providing emotional support within collaborative processes. Specifically, they expected GenAI to analyze communication patterns between narrative writers and visual designers, summarize frequently asked questions or areas needing clarification, and facilitate smoother interactions. W6 suggested, \textit{“If designers frequently ask questions to AI, it could analyze their queries and responses, summarizing parts that were difficult to understand or required clarification. This would help me understand designers' challenges better and identify areas where I can improve communication to enhance collaboration."}

\re{Participants described GenAI as a neutral communication intermediary that could lower the psychological barrier to initiating communication, particularly in situations where creators hesitated to ask questions or raise concerns during email or chat-based collaboration. By relaying requirements clearly, adjusting message tone, and clarifying unfamiliar terms, GenAI was seen as a way to support communication without adding interpersonal pressure.}

\re{Beyond communication mediation, participants also highlighted GenAI’s potential as an emotional companion, offering psychological support that alleviates stress, loneliness, and interpersonal tensions frequently encountered during intensive collaborative work. Narrative writers and visual designers proposed that GenAI could adopt an approachable conversational persona—encouraging, humorous, and empathetic—to improve interpersonal dynamics, boost team morale, and ultimately enhance overall creative productivity.}

\end{itemize}
\subsubsection{Challenges in Visual Concept Art Creation Phase} {\textit{ }}
{\textit{\textbf{The Dilemma of Worldbuilding Accuracy: The Endless Learning Journey of Visual Designers.}}}
\re{Previous studies suggest that visual designers often spend considerable time searching for terms used by writers \cite{cornish2015visual}. Our study extends this finding by showing that designers go beyond simple term searches, investing substantial effort in understanding narrative contexts and historical details to accurately represent the world. This highlights an asymmetry of knowledge and expertise: while narrative writers construct worlds rich in historical and cultural references, visual designers must undertake additional learning to visualize them faithfully. D2 noted, \textit{“I have learned about fantasy universes by watching YouTube videos and movies.”} Similarly, D6 described needing to research not only how medieval armor was worn but also the intent behind its design. Despite these efforts, designers still worry about maintaining historical accuracy, which they view as essential for producing high-quality games and establishing a solid foundation for gameplay.}

{\textit{\textbf{Drafting Under Pressure: Balancing Creativity, Writer's Intent, and the Pitfall of Generic Designs.}}}
\re{In the visual concept art creation and communication phase, visual designers’ creativity is often constrained by competing demands: they must faithfully reflect the writer’s intent and maintain historical accuracy, while also producing compelling and original visuals that define the game’s identity. This tension reflects an asymmetry of goals, as writers prioritize precise realization of narrative intent, whereas designers emphasize creativity and aesthetic appeal. Under tight deadlines and limited access to inspiring references, designers are often forced to compromise by producing safe and generic drafts rather than exploring more innovative options. Visual designers typically present multiple drafts to align on creative intent; as D3 noted, “The more drafts we prepare, the higher the chance one will resonate with the writers.” However, W5 and W7 emphasized that establishing a clear direction early in the drafting process often leads to better outcomes.}

{\textit{\textbf{Compromising Creative Quality under Time Constraints.}}} \re{Tight deadlines significantly constrain the quality of concept art, particularly in character design, which plays a central role in a game’s aesthetics and commercial success. Given the substantial time required for visual work, designers often have little choice but to compromise quality or forego iterative revisions. Narrative writers, aware of these constraints, likewise tend to minimize or avoid requesting changes. As a result, collaboration across roles deteriorates and the quality of final outputs suffers. Time pressure thus acts as a catalyst that amplifies existing asymmetries: while common ground is typically built through sustained communication and iterative feedback, limited time undermines these processes and prevents effective alignment between designers and writers. Although AI-based image generation tools have been explored as potential solutions, existing tools neither adequately support designers’ expertise nor meaningfully improve workflow efficiency in practice.}

\subsubsection{Design Suggestions}{\textit{ }}
\begin{itemize}
\item{\textit{\textbf{Bridging Historical Knowledge Gaps through GenAI-Provided Contextual References.}}} \re{In the process of understanding game narrative documents to create visual concept art, visual designers emphasized the need for GenAI tools that can facilitate easier learning of historical contexts and ensure accuracy when creating visual concept art based on game narrative documents. Designers D2 and D6 particularly highlighted expectations for functionality that instantly provides historically accurate information relevant to their specific tasks. This is because, especially in the initial stages, designers often struggle to determine what they even need to know, resulting in significant time spent grasping the overall context and establishing directions for exploration. Designer D6 explained, \textit{“Traditional historical references are typically text-based, making it difficult to visualize or understand historical backgrounds visually. If GenAI could retrieve and provide not only textual information but also related images or video materials, it would greatly simplify learning historical contexts and effectively visualizing them.”} Similarly, designer D2 proposed, \textit{“We need a GenAI feature capable of reviewing and enhancing historical accuracy during the initial design drafting process.” }He further elaborated, \textit{“For instance, if the AI could recognize the historical context relevant to my current task and immediately display historical reference images or design insights directly related to my ongoing work, it would help quickly identify and correct any historically inaccurate elements in my designs.”} By making historically grounded context visible at the moment of design, such GenAI support could shift historical accuracy from an individual learning burden to a shared, workflow-integrated resource.}

\item{\textit{\textbf{Accelerating Idea Alignment through Rapid Generating.}}} Visual designers wish to utilize GenAI to rapidly prototype initial drafts when creating visual concept art. The purpose of creating these initial drafts is to facilitate idea sharing between writers and designers, align their thought processes, and generate a variety of concepts reflecting the writer's original intent. For quicker prototyping, designer D4 hoped that GenAI would automatically generate textual prompts based on descriptions within the game narrative documents requiring visualization. While writers preferred GenAI to generate images directly, designers wanted to maintain control over image creation, opting instead to personally manage text prompts. Designers also expressed a desire for GenAI to produce multiple variations of images based on the same prompt, thereby increasing their creative options. Designer D6 anticipated that GenAI could significantly shorten the concept art review process. He expected that when writers provide feedback on initial concepts, designers could quickly implement rough adjustments, after which GenAI could immediately produce high-quality finalizations, thus minimizing repeated rounds of feedback.

\item{\textit{\textbf{Customizing AI Training to Match Creative Vision and Style.}}}
Visual designers desire GenAI tools optimized specifically for their workflows and working styles. D4, who has extensive experience with various AI tools, emphasized that most existing tools are primarily focused on text-to-image generation, while highlighting a greater need for image-to-image functionality that refines existing artwork. For instance, even a single character design requires additional variations such as diverse facial expressions, poses, and outfits. If GenAI could efficiently assist in image modification, it would significantly reduce production time. Recently, studies have begun exploring image-to-image techniques to create game design assets. Designers D1 and D6 expressed a strong preference for tools capable of directly learning the game's unique art style and generating results based on their own sketches. They stressed the importance of directly managing the AI model training process, stating that having control over this process is crucial to aligning AI-generated outputs with their creative visions. Such direct management allows designers to evaluate and adjust learning outcomes in real-time without repeatedly depending on external AI teams. Most importantly, customized tools like these would substantially reduce repetitive tasks and provide designers with greater time flexibility, thus enabling them to focus more effectively on communication and collaboration with narrative writers.
\end{itemize}

\begin{table}[t]
    \scriptsize
    \centering
    \setlength{\tabcolsep}{4pt}
    \renewcommand{\arraystretch}{1.1}

    \begin{tabularx}{\columnwidth}{
        >{\raggedright\arraybackslash}m{2.3cm}
        >{\raggedright\arraybackslash}X
        >{\raggedright\arraybackslash}X
    }
        \toprule
        \textbf{Workflow stage} 
        & \textbf{(i) Collaboration challenges} 
        & \textbf{(ii) Envisioned GenAI mediation roles} \\
        \midrule

        \makecell[vl]{\rule{0pt}{6ex}Narrative\\construction} 
        &
        \begin{itemize}[leftmargin=*, topsep=0pt, itemsep=2pt, parsep=0pt]
            \item Early breakdown in stabilizing, structuring, and externalizing narrative ideas into shareable forms prior to collaboration
            \item Writers’ reliance on individual memory leads to narrative information loss, causing misalignment and inconsistencies across collaboration.
        \end{itemize}
        &
        \begin{itemize}[leftmargin=*, topsep=0pt, itemsep=2pt, parsep=0pt]
            \item GenAI provides genre-aligned grounding that helps transform abstract early ideas into shareable starting points for collaboration
            \item GenAI lowers the documentation barrier by turning spontaneous conversational ideas into minimally structured, shareable drafts
            \item GenAI preserves narrative knowledge and validates consistency to serve as a shared memory for continuous alignment
        \end{itemize}
        \\

        \midrule
        \makecell[vl]{\rule{0pt}{6ex}Narrative document \\communication} 
        &
        \begin{itemize}[leftmargin=*, topsep=0pt, itemsep=2pt, parsep=0pt]
            \item Narrative documents lack sufficient context and knowledge for visualization, leading to misalignment of intent.
            \item Hierarchical dynamics and differing work styles discourage questions and clarification.
            \item Asynchronous communication delays feedback, forcing work to proceed with incomplete alignment.
        \end{itemize}
        &
        \begin{itemize}[leftmargin=*, topsep=0pt, itemsep=2pt, parsep=0pt]
            \item GenAI transforms text-based narratives into visual references, enabling clearer sharing of authorial intent with designers.
            \item GenAI reduces knowledge gaps and time–space constraints through real-time access to and Q\&A over narrative documents.
            \item GenAI mediates communication and emotional tensions, reducing misunderstandings and supporting smoother collaboration.
        \end{itemize}
        \\

        \midrule
        \makecell[vl]{\rule{0pt}{6ex}Visual concept \\ art creation}
        &
        \begin{itemize}[leftmargin=*, topsep=0pt, itemsep=2pt, parsep=0pt]
            \item Visual designers must independently learn complex historical and narrative contexts to ensure accuracy, creating knowledge asymmetry and delaying alignment with writers.
            \item Under time pressure, designers lack efficient ways to rapidly prototype and iterate on visual concepts, slowing idea alignment with writers.
            \item Generic AI tools fail to reflect project-specific styles and workflows, forcing designers to spend time on repetitive refinements instead of collaboration.
        \end{itemize}
        &
        \begin{itemize}[leftmargin=*, topsep=0pt, itemsep=2pt, parsep=0pt]
            \item GenAI provides task-relevant historical context and visual references on demand, reducing learning overhead and supporting early alignment in visual design.
            \item GenAI enables rapid generation of multiple visual variations from narrative descriptions, accelerating early prototyping and cross-role idea alignment.
            \item GenAI supports designer-controlled, style-specific training and image-to-image refinement, reducing repetitive labor and freeing time for collaborative work.
        \end{itemize}
        \\
        \bottomrule
        \addlinespace[5pt]
        \textbf{(iii) Creators’ Concerns about Collaboration with AI}
        &
        \multicolumn{2}{>{\raggedright\arraybackslash}m{1\linewidth}}{
        \begin{itemize}[leftmargin=*, topsep=2pt, itemsep=2pt, parsep=0pt]
            \item Concerns about visible AI use undermining professional competence and accountability.
            \item Risks of eroded collaborative trust and accountability conflicts due to inaccurate or misaligned AI outputs.
            \item Increased learning and skill burdens associated with adapting to GenAI workflows.
            \item Concerns about AI’s impact on long-term careers, creative identity, and entry opportunities.
            \item Risks of reduced direct communication under efficiency-driven AI-mediated workflows.
        \end{itemize}
        }
        \\
    \bottomrule
    \end{tabularx}

    \caption{\re{Synthesis of findings showing (i) stage-specific collaboration challenges, (ii) GenAI mediation roles proposed to address these breakdowns, and (iii) cross-cutting concerns creators express about collaborating with GenAI.}}
    \label{tab:themes_summary_genai}
\end{table}

\section{\MakeUppercase{Creators' Concerns About Collaboration with AI}}
\label{sec:concerns}
Creators are not always enthusiastic about integrating GenAI to their creative process. From hearing their perspectives, we learned that there are concerns about collaborating with GenAI. Creators do not necessarily view GenAI positively, and there are several psychological barriers that need to be addressed before they can truly recognize AI as a collaborator.

\textit{\textbf{Concerns about Appearing Less Competent:}}
Our study explored how creators perceive AI use within collaborations among fellow creators, including their preferences regarding transparency of AI involvement. Participants expressed particular concern that revealing their real-time collaboration process with AI could negatively impact their professional image. Consequently, they preferred sharing only the final outputs rather than the entire collaborative process. Participant W6 explained, \textit{“When sharing work with the design team, I usually only show the final results, not the process. There is still prejudice against using AI, and I don't see much benefit in exposing the trial-and-error involved in creating a story. It feels more appropriate to summarize or organize the AI conversations before sharing.”} Similarly, participant D5 added, \textit{“Art is a process of continuously developing one’s individuality and style over time. However, the perception that results can be produced simply by clicking a button remains negatively viewed by other artists and creators.”} Additionally, participants expressed concerns that errors or mistakes caused by AI would ultimately reflect on them, potentially damaging their professional reputation. Even if AI generates or significantly contributes to the creative work, creators felt they would still be held responsible for any inaccuracies. This situation places significant pressure on creators, as it could lead to negative evaluations of their professional competence.

\textit{\textbf{Concerns about Eroding Collaborative Trust among Creators:}}
Participants emphasized that the accuracy and reliability of information and visual materials provided by AI are critical issues in collaborative creative work. They noted that when AI provides inaccurate or contextually incorrect data related to story settings or character details, it can lead to confusion and trust issues among creators. If creators uncritically accept AI-generated materials, incorrect information may become embedded in the final outputs, potentially causing conflicts over accountability or reducing overall trust in the AI itself. Participant W2 explained, \textit{“When AI provides incorrect outcomes, the responsibility ultimately falls on me. Therefore, rather than blindly trusting AI outputs, I always need to double-check them."} Similarly, D4 expressed, \textit{“Honestly, incorrect information from AI can cause significant confusion. It can even steer the entire team in the wrong direction, so I always remain cautious when using AI-provided materials."} Consequently, establishing a rapid verification process for AI-generated content, along with transparent systems that clearly reveal the sources of AI materials, is essential. These measures can enhance trust among creators, improve collaborative efficiency, and ultimately contribute to the quality of creative outcomes.

\textit{\textbf{Concerns about the Burden of Adapting to GenAI in Creative Collaboration:}}
Creators feel pressured to adapt to AI in the creative process to avoid obsolescence. D6 noted that AI adoption adds the burden of learning new skills alongside creating art. This pressure is intensified by reduced job opportunities, forcing creators to take on multiple roles. W2 highlighted that prompt-writing skills in AI collaborations are overshadowing traditional skills like sketching. D4 and W4 emphasized that proficiency with AI tools significantly impacts the quality and diversity of results, making mastery of these tools an increasingly heavy burden for creators.

\textit{\textbf{Concerns about AI Threatening Creators' Careers and Growth:}}
Creators perceive AI as both a tool and a competitor. W2 expressed concerns that growing reliance on AI could diminish creators' unique roles, leading to feelings of defeat. D1 shared similar worries, fearing that AI might replace their creative work, resulting in a sense of loss. Additionally, AI is seen as a barrier for beginners, with D6 warning that it could eliminate opportunities for novice designers to start their careers. D1 emphasized that the next generation of creators might shrink as AI outpaces human skill development.

\textit{\textbf{Concerns about Losing Collaborative Communication due to AI:}}
A significant concern that emerged from these discussions is the potential for AI tools to hinder the importance of communication. D4 emphasized that while AI tools can enhance convenience and efficiency, it is crucial not to overlook collaboration and communication. They stressed the importance of maintaining regular meetings and feedback sessions to ensure consistent communication, even when integrating AI tools into practical workflows. D4 explained, \textit{“While these tools can help save time on tasks, they shouldn’t lead to a lack of communication. The tools are there to make work easier, but that doesn’t mean the emphasis on communication should decrease.”} They highlighted the need to use AI tools as a means to streamline work while preserving open and active collaboration.

\section{\MakeUppercase{Discussion}}

\subsection{Strategies for Positioning GenAI as a Trusted Creative Collaborator}
\label{dis:one}
This section addresses ways to improve collaboration between creators using GenAI tools and help creators perceive GenAI as a better partner. While previous studies emphasize the need to assist creators in adopting GenAI positively, there is a lack of empirical research on specific concerns and how to address them. Our study identifies the factors that hinder creators from accepting GenAI as a collaborator and suggests ways to resolve these issues so that GenAI is perceived as a tool that complements the creativity and capabilities of creators.

\textit{\textbf{Personalized GenAI Companions for Protecting Creative Agency in Collaboration.}}
\re{Creators expressed a desire to use personalized GenAI “companions” to protect their creative work and ensure that their expertise is respected during collaboration with GenAI. Such systems are particularly valuable when working with sensitive information or undisclosed projects, as they allow creators to selectively share ideas and work outputs. For example, GenAI could be configured to share only specific portions of work between collaborators. In a writer–designer collaboration, the designer might refine a character’s visual design with GenAI, informed by discussions with the writer, while choosing to share only selected design drafts and withholding early sketches or experimental iterations. Beyond information protection, these practices reflect a broader desire to preserve creative agency and emergent creativity in GenAI-mediated collaboration. We therefore caution against overly intrusive GenAI intervention in collaborative creative processes. Prior work has shown that ambiguity and partial understanding, rather than complete shared understanding, can foster new interpretations and emergent outcomes. This suggests that excessively high levels of coordination may paradoxically limit opportunities for unexpected ideas or novel interpretations \cite{ladenheim2021babyface, guridi2025fake}. Accordingly, future research should explore how GenAI can support shared understanding without undermining the emergent processes central to creative collaboration, and how to strike an appropriate balance between coordination and creative openness.}

\textit{\textbf{Designing GenAI as a Resilient Facilitator in Creative Collaboration.}} 
\re{Prior research \cite{wang2024ai, chen2024persona, tang2023enhancing, amin2025generative} has primarily emphasized the importance of persona customization in interactions with GenAI within individual work contexts. In contrast, our study foregrounds the distinct persona characteristics required in collaborative work settings. Our findings show that, in collaborative contexts, participants envisioned GenAI as a facilitator that supports communication and mutual understanding among team members. This collaborative persona was expected to provide clear and precise information, reduce misunderstandings, and enable smooth interaction. Participants particularly emphasized the importance of a GenAI persona that is formal yet friendly, maintaining neutrality while also demonstrating empathetic and attentive behavior. Furthermore, participants expressed concerns that inaccurate GenAI outputs, when shared among collaborators, could amplify confusion and undermine trust. To mitigate these risks, we propose designing AI-resilient mechanisms that support users in understanding the limitations, uncertainty, and potential errors of GenAI outputs within creative collaboration contexts \cite{glassman2024ai}. While prior work on AI-resilient mechanisms has largely focused on individual user–AI interactions \cite{gu2024ai}, we argue for the need to extend these discussions to collaborative settings. In such contexts, GenAI mediates interaction among participants and functions as a boundary object that connects diverse perspectives and contextual knowledge. At the same time, GenAI can mediate AI-resilient mechanisms that help creators avoid overreliance on AI outputs and support collective recognition and response when errors or potential harms are identified.}

\textit{\textbf{Promoting Communication and Collaboration with GenAI Tools.}}
To prevent GenAI tools from causing a reduction in communication among creators, GenAI should be utilized in ways that supplement and encourage human collaboration and communication. Even when AI handles automated tasks, it is crucial for team members to jointly review and discuss GenAI-generated outcomes, thereby enhancing trust and interaction. Recent studies \cite{marrone2022creativity, al2024balancing, haase2024human, hosanagar2025designinghumangenerativeai} emphasize that AI-generated ideas should pass through a human judgment ‘filter' during the creative process, highlighting the importance of human trust and control, particularly when AI serves collaborative roles. While these studies primarily focused on individual interactions between humans and GenAI, our research uniquely emphasizes that maintaining human agency is even more critical within collaborative contexts involving multiple creators. We highlight that collective human judgment, communication, and decision-making become especially significant when GenAI tools mediate interactions among creators. For example, when Creator A sends a message to Creator B through GenAI, the system could actively promote additional communication by prompting, “GenAI might make mistakes. Would you like to directly confirm accuracy with Creator B?" In this way, GenAI can serve not merely as a tool in the creative process but also as a facilitator that actively enhances human collaboration and communication.

\textbf{\textit{Supporting Creators' Emotional Well-being and Collaborative Harmony.}}
In collaborative relationships, creators perceived AI not merely as a task assistant but rather as an emotional companion that supports their emotional stability and psychological well-being. For instance, AI could function as an emotional curator, sensing creators' psychological states (e.g., stress or reduced confidence) and assisting them in maintaining an optimal creative mindset. GenAI could recognize moments of stress and help creators regain emotional balance by reminding them of past achievements or highlighting positive feedback they've previously received. Additionally, during collaborative processes, when uncomfortable emotions or tensions arise between creators, GenAI could act as a neutral and amicable mediator, helping resolve misunderstandings, facilitating smoother communication, and alleviating interpersonal conflicts.

\subsection{Design Implications for GenAI: Enhancing Collaboration in Creative Teams}
\label{dis:two}
In a prior study, AI was perceived as a central component of a collaborative ecosystem where users, technologies, and organizational structures interact \cite{10.1145/3613904.3642697}. Our study finds that GenAI is increasingly positioned not only as a tool for producing creative outputs, but also as a mediator that enhances interaction between creators.

Through our study, we identified three key functions that GenAI could perform to enable more effective collaboration between creators from different domains:
(1) supporting creative workflows to enhance communication opportunities,
(2) augmenting missing or unclear information in real time, and
(3) providing psychological stability by anticipating and resolving potential conflicts.

\re{First, participants expected GenAI to flexibly switch between an assistant role for individual creators and a mediator role for inter-creator collaboration. We envision a system in which GenAI operates across two interconnected spaces: a private space for individual ideation and a shared space for collaboration. In the private space, creators work one-on-one with GenAI while the system accumulates contextual data to understand creators' intentions. In the shared space, this context enables collaboration—for example, allowing a visual designer to query a narrative writer's creative intent. GenAI mediates semantic gaps by translating individualized creative intent into shared understanding.}

\re{Second, GenAI can detect and clarify missing or ambiguous information that commonly arises between creators from different disciplines. Interviews revealed that even when documents are shared, background details or visual cues are often underspecified, leading to confusion or misinterpretation. In such cases, GenAI could proactively prompt clarification, for example: “This scene lacks specific visual details. Would you like to consider referencing 19th-century French urban architecture?” By identifying and resolving underspecified content, GenAI functions as an active information facilitator that supports accurate interpretation and reduces misunderstanding.}

\re{Third, GenAI can help maintain psychological safety and mitigate conflict by identifying early signs of tension during collaboration. Participants described challenges stemming from hierarchical directives, hesitation to challenge others’ expertise, or mismatched communication styles. As a neutral communication buffer, GenAI can prompt creators to provide additional context, invite feedback, or notify collaborators of impactful changes (e.g., “This request may conflict with the current visual direction. Would you like to provide additional context or invite feedback?”). In doing so, GenAI helps alleviate interpersonal strain and fosters a more emotionally supportive collaborative environment.}

Based on this insight, we propose a user interface concept we term “Dual-Space GenAI.” In this model, the GenAI platform features two distinct yet connected modes: a Personal Studio and a Shared Workshop.

\re{Personal Studio is a private, sandboxed environment where creators interact one-on-one with GenAI. A key feature is the Contextual Memory toggle, which, when enabled, allows GenAI to record creators’ decision-making processes, discarded ideas, and stylistic preferences as structured metadata. For example, if a writer explores multiple character backstories before selecting one, the GenAI logs this process. Prior to sharing, GenAI can also proactively identify potentially ambiguous expressions and suggest refinements into clearer, collaboration-friendly language to reduce misunderstanding. Shared Workshop is the collaborative space. When a creator shares an artifact (e.g., a story draft) from the Studio to the Workshop, they are prompted with Granular Context-Sharing Controls, enabling them to share not only the final artifact but also selected elements of the associated Contextual Memory. A collaborator can then query the GenAI about prior creative considerations (e.g., alternative backstories), and the system can respond by explaining the creator’s reasoning. This approach moves beyond simple file sharing toward the mediated transfer of creative intent, directly addressing the semantic gaps identified in our findings.}

\re{To realize the Dual-Space GenAI concept, future research must address challenges on two fronts: core system implementation and effective creator use. On the system side, this includes structuring nuanced, unstructured creative-process data into a queryable knowledge representation, paired with robust protections for privacy and intellectual property. On the interaction side, it requires designing intuitive interfaces that provide granular control over sharing sensitive contextual information, while developing non-intrusive AI mediation models that enhance collaboration without disrupting creative flow.}

\re{Our study derives these insights from the game development domain by examining collaboration between heterogeneous creators with distinct roles, namely designers and writers. While grounded in this specific context, our findings are not domain-bound; rather, they offer broader implications for the CSCW community concerned with AI-mediated collaboration. In particular, this work systematically identifies concrete needs and design implications for GenAI systems that support collaboration among heterogeneous creators working across different modalities and areas of expertise. Beyond the game creation context, our discussion reframes GenAI as an active collaborative mediator, opening up new avenues for future CSCW research to examine how GenAI systems can support sensemaking, coordination, and trust in heterogeneous and cross-disciplinary collaborative settings.}

\subsection{Takeaways from a Participatory Design Study Using Creative Prompting}
In this study, we employed the Participatory Design approach, widely recognized in the HCI community \cite{zhang2022algorithmic, 10.1145/3544548.3581386, kim2024unlocking}, to collaboratively design GenAI concepts supporting effective collaboration among creators with different expertise. Here, we highlight key insights from our participatory design study to inform future research and practice on GenAI-supported collaborations.

Through initial observations and interviews with GenAI prompting, participants concretely understood how GenAI could impact their collaborative work. We found preliminary interviews or literature reviews essential for participants to accurately contextualize their collaborative challenges, enhancing their engagement during subsequent exploration activities.

Structured missions, guided by insights from preliminary interviews, allowed participants unfamiliar with GenAI to directly envision practical enhancements to their collaborative processes. Hands-on experiences through simple prompting activities and immediate feedback from GenAI significantly deepened their understanding, inspiring innovative ideas previously considered abstract or unattainable.

Effective participant engagement with GenAI required thorough preparation and proactive facilitator support. Although pre-session educational materials (e.g., instructional videos) were provided, facilitators’ active real-time assistance, such as answering questions, guiding demonstrations, and encouraging screen-sharing, was indispensable for building participants' confidence and proficiency within limited session times.

Additionally, providing diverse and stimulating materials enriched participants' creative ideation. We developed generative AI function cards, practical intervention examples, and concise idea summary sheets. Function cards specifically enabled participants unfamiliar with GenAI to easily conceptualize features using everyday language and visuals. Introducing advanced GenAI capabilities beyond the tools directly used by participants (e.g., ChatGPT, Midjourney) further expanded their creative explorations, prompting broader and more innovative ideas.

Finally, clearly structured guidelines helped participants balance the strengths and potential concerns of GenAI-based solutions. Encouraging comprehensive reflection on both benefits and ethical or practical limitations facilitated realistic and practical outcomes. Creating an empathetic environment where participants openly discussed real-world collaboration challenges was also essential in identifying meaningful opportunities for GenAI support.

Collectively, these methodological and practical insights confirmed the effectiveness of co-design sessions for domain experts to generate realistic and practical GenAI solutions. These findings can broadly inform Participatory Design (PD) processes involving diverse creator groups, such as engineers, designers, and artists, fostering deeper understanding of their collaborative challenges and potential technology-supported solutions. Ultimately, collaboration tools bridging creators and GenAI can significantly enhance creative effectiveness across diverse creative domains.
\section{\MakeUppercase{Limitations}}
We acknowledge several limitations in this study. First, as the research focused on game writers and designers in Korea, the findings may not fully represent the broader population or creators from other regions, whose characteristics and practices might differ. Additionally, the participants primarily consisted of designers and writers aged 20 to 50, with no representation from creators over 60. This demographic limitation may have influenced the results, particularly as the materials provided could present challenges for those with limited technological literacy. Future research will aim to address these limitations by including creators from diverse cultural backgrounds and age groups, as well as those with minimal prior experience with AI tools, to develop more inclusive and comprehensive solutions.

Second, the generative AI capability cards and examples of interventions presented to the participants in this study may have been limited to what currently exists. While the study provided guidance to encourage more creative thinking beyond the AI capability cards and interventions, we also used ChatGPT-4 and MidJourney in the participants' prompt-writing process. Although these tools represent some of the best technology we are aware of, AI is continuously evolving, and new technologies are being introduced. Therefore, ongoing research into AI stimuli that adapt to these changes will be necessary.

Third, this study focused on single sessions with individual participants within a short timeframe, allowing for an in-depth exploration of their experiences and insights. However, the breadth of insights that could be gained from iterative idea development over an extended timescale, involving multiple sessions and diverse stakeholders, may have been limited. Drawing on previous studies that observed creator-AI interactions over longer periods \cite{10.1145/3643834.3661587}, future research plans to expand the scope by incorporating multi-session co-design approaches with a variety of stakeholders, aiming to derive richer and more comprehensive insights.

Fourth, this study focused on the remote collaboration process between two creators, as pre-study findings indicated that this approach was more commonly used and uncovered more challenges. While the study provided valuable insights into how GenAI can support this process, it may not have fully addressed the collaborative dynamics and challenges of creators working with AI in a shared workspace. Future research is needed to explore emerging trends and tackle the unique challenges of shared work environments.

Lastly, we found that some challenges, such as deep organizational issues or interpersonal conflicts, may be difficult to resolve solely through GenAI and may require human-centered interventions. However, this study focused on exploring the possibilities and opportunities of GenAI, with an emphasis on identifying how GenAI can be applied to enhance collaboration. Through this approach, we uncovered a wide range of promising opportunities for GenAI to facilitate future collaborative processes. That said, it is important to remain cautious of AI universalism, recognizing that not all problems can be solved by GenAI alone. For challenges that require non-GenAI approaches, we plan to address them in future research endeavors.

\section{\MakeUppercase{Conclusion}}
Our study demonstrates that Generative AI (GenAI) tools have great potential not only to support individual creative work but also to enhance collaboration between creators. We conducted preliminary interviews with game narrative writers and visual designers to understand their collaboration processes, and then applied a multi-method approach combining semi-structured interviews and a participatory design study. Through the semi-structured interviews, we identified multi-dimensional challenges and asymmetries that span the entire creation workflow, encompassing from preparation phases to interactive co-creation processes of working together. To address these, we facilitated the participatory design study. During these sessions, we employed creative prompting and various stimuli based on the work processes of narrative writers and visual designers to derive design solutions tailored to their needs. Through the design sessions, we found that GenAI tools can help creators manage their stress with emotional connection and organize ideas through a vocal assistant, anytime and anywhere. Additionally, we found that GenAI has the potential to enhance real-time communication by bridging knowledge and workflow gaps between collaborators, streamlining the feedback process through rapid draft creation, and reducing the psychological burden associated with collaborative interactions.

In contrast, our research also revealed concerns about the use of AI in collaboration. Beyond practical worries such as privacy and reduced communication, participants raised relational and procedural concerns about how GenAI might reshape interpersonal dynamics and decision-making in creative processes. We emphasize that when designing GenAI tools, it is crucial to provide flexible personalization options for AI personas, protect creators' privacy during collaboration, and actively foster communication among creators while ensuring transparency and accountability. In conclusion, while GenAI tools hold significant potential to enhance collaboration between creators, careful design is necessary to realize this potential. In summary, this study contributes to a deeper understanding of the challenges underlying heterogeneous creative collaboration and identifies design opportunities for GenAI to function as a facilitator and mediator that supports—rather than replaces, human creativity and values. At the same time, our findings highlight critical concerns that must be addressed in future system design, offering balanced guidance for developing GenAI tools that can truly strengthen creative collaboration.

\begin{acks}
To Robert, for the bagels and explaining CMYK and color spaces.
\end{acks}

\bibliographystyle{ACM-Reference-Format}
\bibliography{citation}

\newpage
\appendix
\section{Creative-Prompting Session Example Prompts}
\label{appendix:exampleprompts}

Example prompts are provided for each step. Overly long or uninformative AI-generated responses were truncated or summarized.  

\enlargethispage{\baselineskip}
\vspace{-0.5\baselineskip}

\begin{figure}[h]
    \centering
    \includegraphics[width=\textwidth]{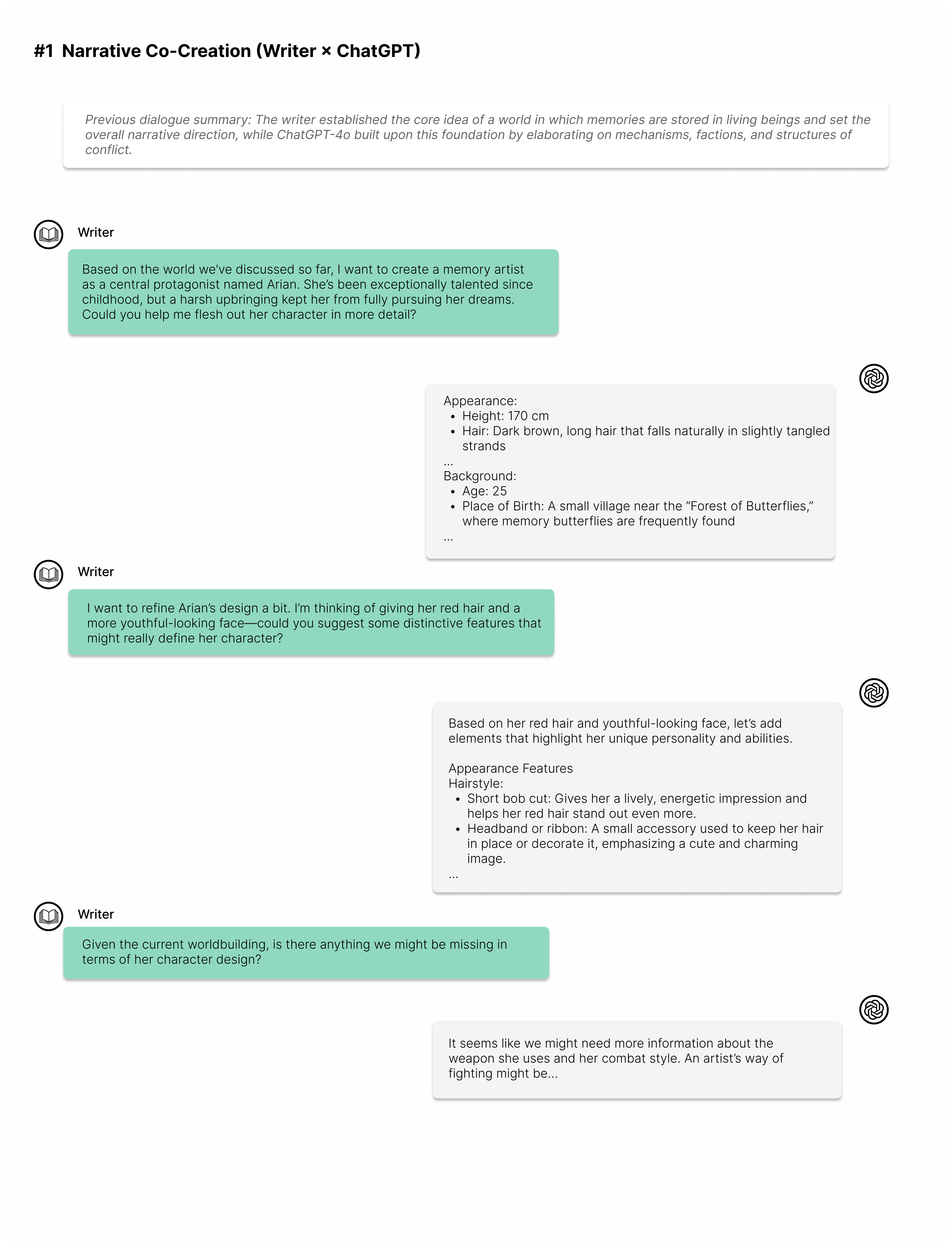} 
    \caption{Creative-prompting session: Narrative Co-creation}
\end{figure}

\begin{figure}[h]
    \centering
    \includegraphics[width=\textwidth]{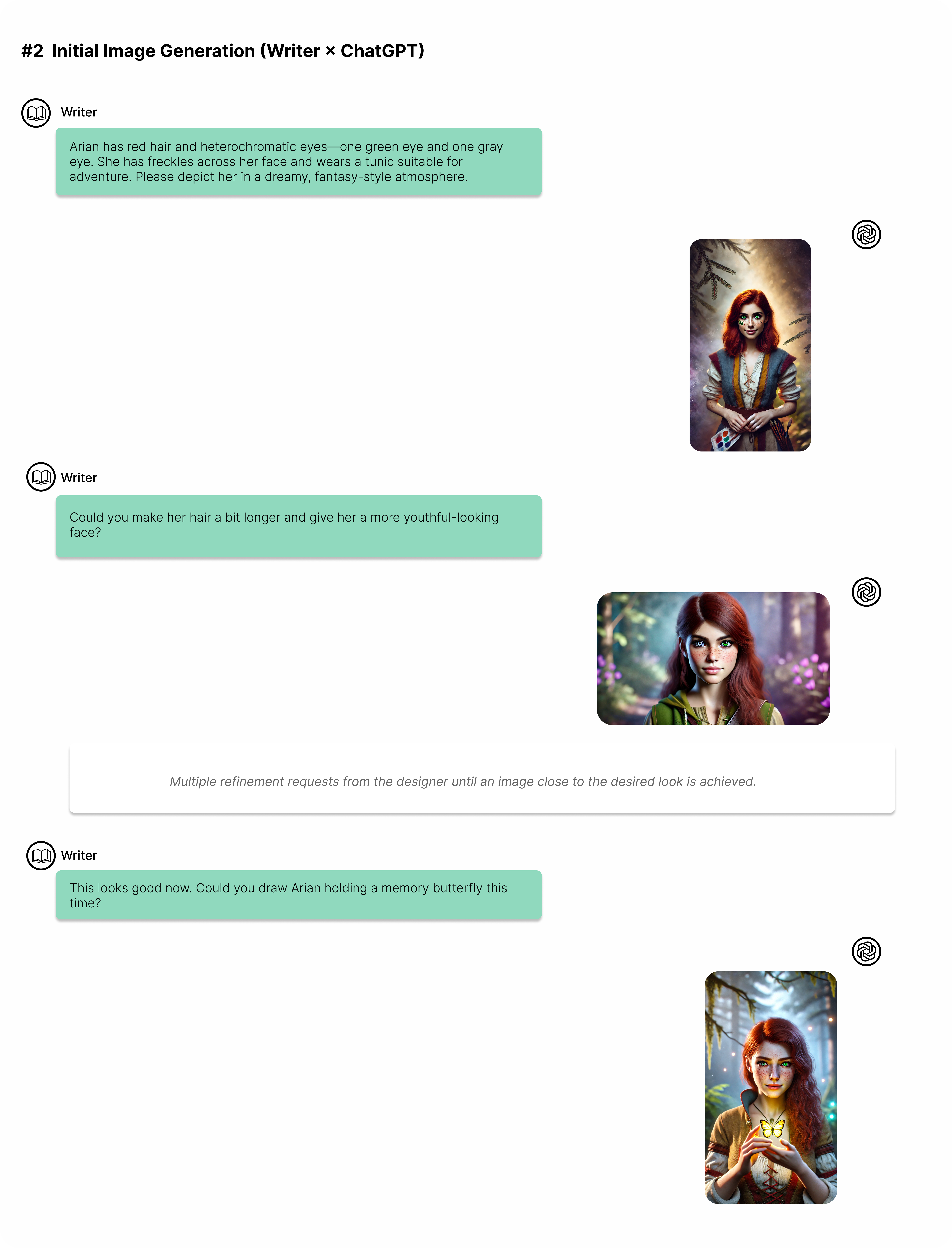} 
    \caption{Creative-prompting session: Initial Image Generation by Narrative Writer}
\end{figure}

\begin{figure}[h]
    \centering
    \includegraphics[width=\textwidth]{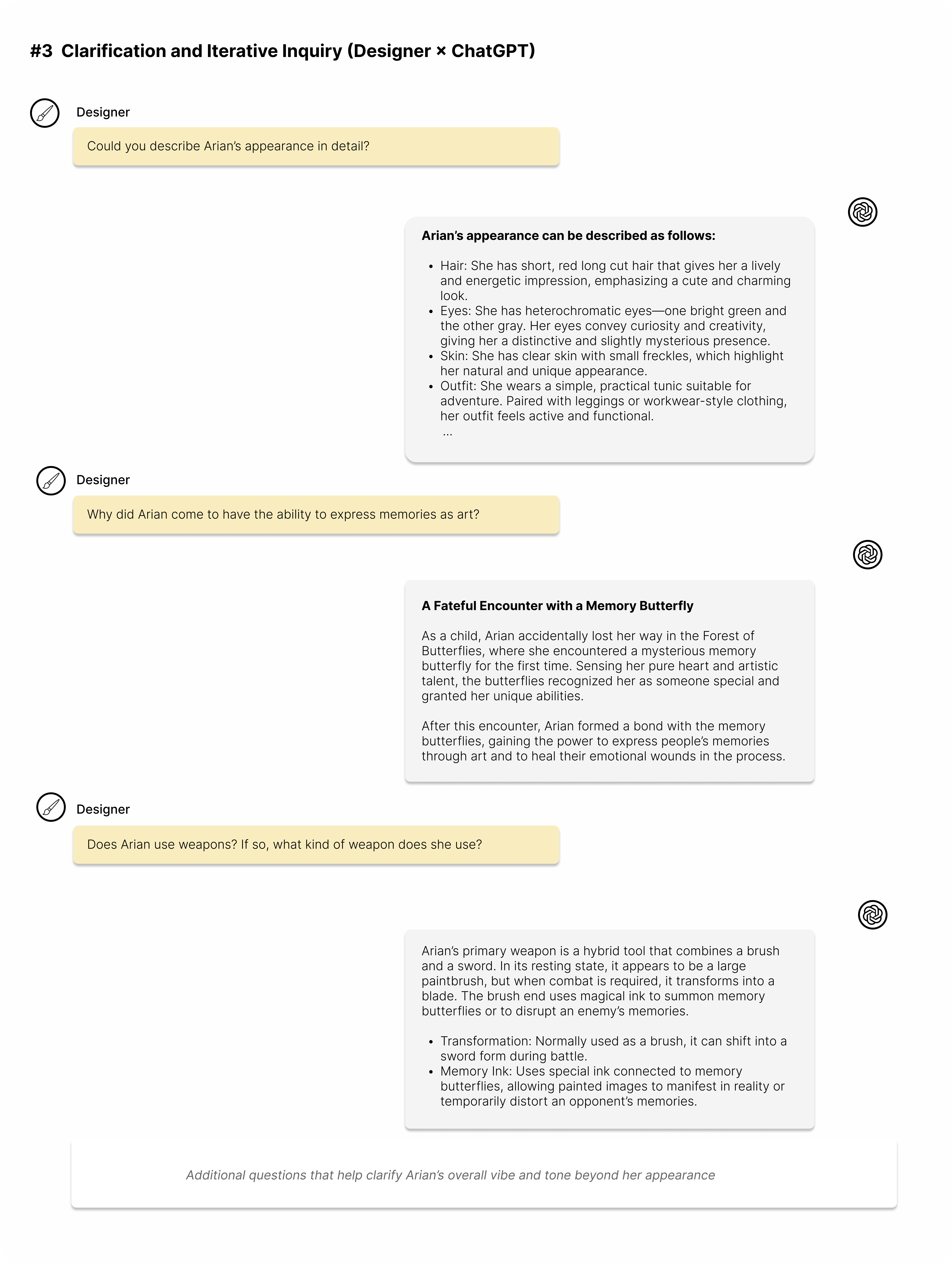} 
    \caption{Creative-prompting session: Clarification and Iterative Inquir}
\end{figure}

\begin{figure}[h]
    \centering
    \includegraphics[width=\textwidth]{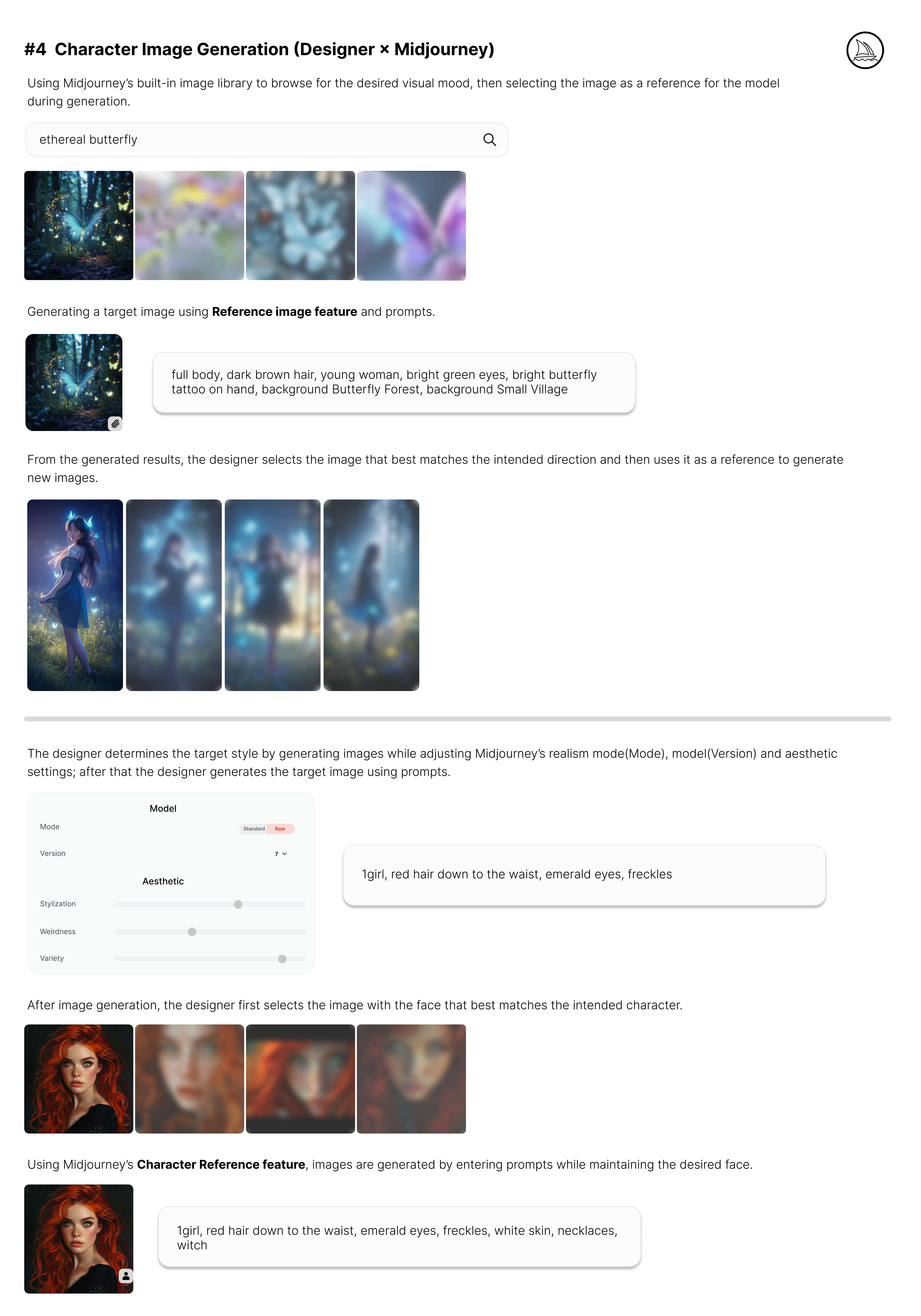} 
\end{figure}

\begin{figure}[h]
    \centering
    \includegraphics[width=\textwidth]{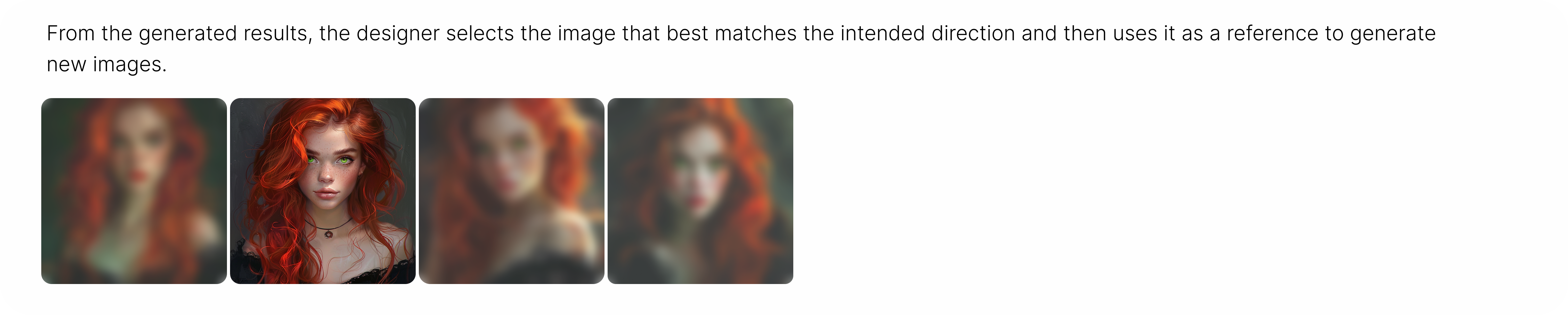} 
    \caption{Creative-prompting session: Character Image Generation by Visual Designer}
\end{figure}

\clearpage
\section{Generative AI Capability Cards}
\label{appendix:capabilitycards}

\begin{figure}[h]
    \centering
    \includegraphics[width=\textwidth]{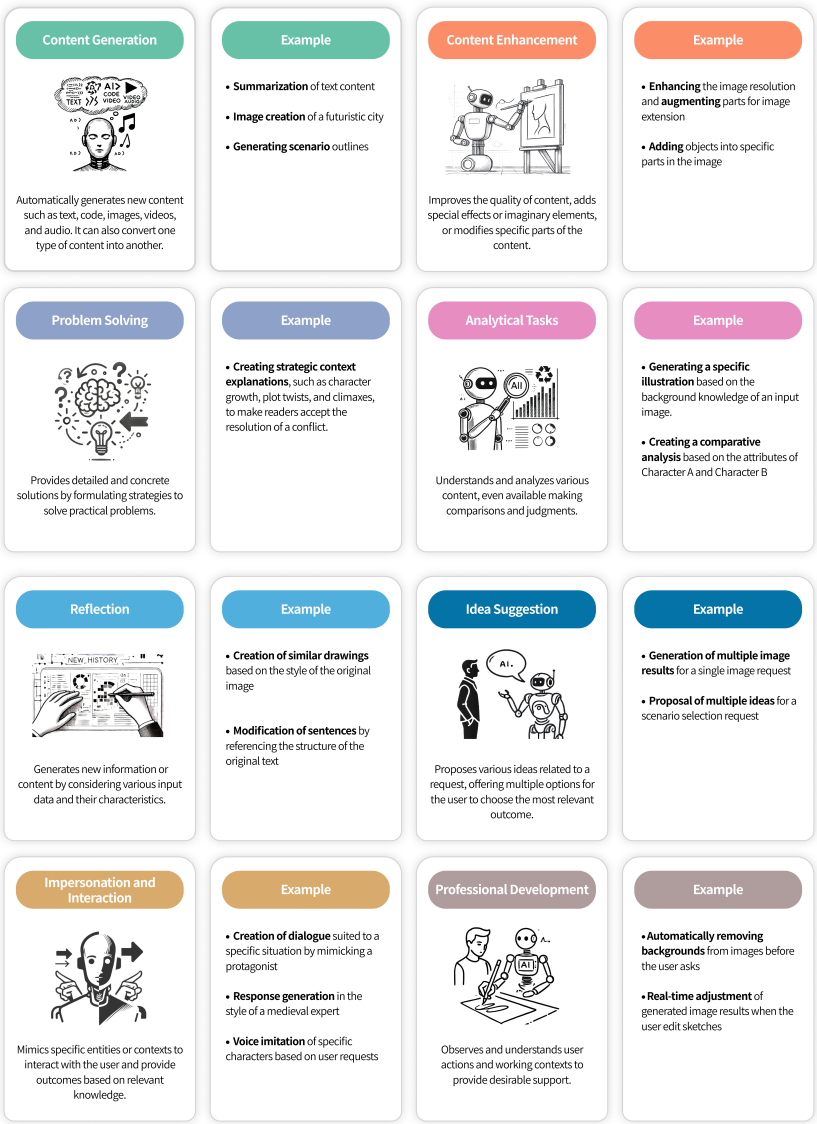} 
    \caption{Generative AI Capability Cards}
\end{figure}

\end{document}